\documentstyle[eqsecnum,aps,prd]{revtex}
\begin{document}
\thispagestyle{empty}
{\baselineskip-4pt
\font\yitp=cmmib10 scaled\magstep2
\font\elevenmib=cmmib10 scaled\magstep1  \skewchar\elevenmib='177
\leftline{\baselineskip20pt\vbox to0pt
   { {\yitp\hbox{Osaka \hspace{1.5mm} University} }
     {\large\sl\hbox{{Theoretical Astrophysics}} }\vss}}

\rightline{\large\baselineskip20pt\rm\vbox to20pt{
\baselineskip14pt
\hbox{HUPD 9604}
\hbox{OU-TAP-34}
\vspace{2mm}
\hbox{\today}\vss}}%
\vskip3cm
\begin{center}{\large
Quantum fluctuations and CMB anisotropies in one-bubble
open inflation models}
\end{center}
\vspace{0.5cm}
\begin{center}
 {Kazuhiro Yamamoto$^1$, Misao Sasaki$^2$ and Takahiro Tanaka$^2$}\\
\vspace{0.5cm}
{\em $~{}^1$ Department of Physics,~Hiroshima University\\
 Higashi-Hiroshima 739,~Japan}\\
\vspace{0.4cm}
{\em $~{}^2$ Department of Earth and Space Science, \\
Osaka University,~Toyonaka 560,~Japan}\\
\vspace{0.4cm}
\end{center}
\begin{abstract}
We first develop a method to calculate a complete set of mode functions
which describe the quantum fluctuations
generated in one-bubble open inflation models. 
We consider two classes of models.
One is a single scalar field model proposed by Bucher, Goldhaber and
Turok and by us as an example of the open inflation scinario,
 and the other is a two-field model such as the ``supernatural'' 
inflation proposed by Linde and Mezhlumian. 
In both cases we assume  
the difference in the vacuum energy density between 
inside and outside the bubble is negligible.
There are two kinds of mode functions. One kind has usual continuous
spectrum and the other has discrete spectrum with characteristic
wavelengths exceeding the spatial curvature scale.
The latter can be further devided into two classes in terms of its origin. 
One is called the de Sitter super-curvature mode, 
which arises due to the global spacetime structure of de Sitter space,
 and the other is due to fluctuations of the bubble wall.
We calculate the spectrum of quantum fluctuations 
in these models 
and evaluate the resulting large angular scale CMB anisotropies.
We find there are ranges of model parameters that are consistent with
observed CMB anisotropies.
\end{abstract}

\section{Introduction}

Motivated by observations that we may live in a low density universe 
\cite{Peebles,RPa,Primack}, several authors have considered 
a possible scenario which realizes an open universe $(\Omega_0<1)$ 
in the context of inflationary cosmology \cite{BGT,YST95,Lindea,Lindeb}.
In contrast to the standard inflation models which
predict a spatially flat universe $(\Omega_0=1)$, 
this scenario predicts an universe with spatially negative curvature. 
In this scenario, the nucleation of a vacuum bubble plays an essential role. 
In general, the bubble nucleation process is described by the 
bounce solution with O(4)-symmetry, which is a non-trivial
classical solution of the field equation in Euclidean 
spacetime\cite{Col,DeLCol}.
Then the expanding bubble after nucleation is described 
by the classical solution obtained by analytic 
continuation of the bounce solution to Lorentzian spacetime.
Owing to the O(4)-symmetry of the bounce solution, the
expanding bubble has the O(3,1)-symmetry. 
This implies that the system is homogeneous and isotropic
on the hyperbolic time slicing inside the bubble and that 
the creation of one
bubble can be regarded as the creation of an open universe.

To obtain a realistic model of an open universe, the flatness
$\Omega_0\sim1$ and the homogeneity and isotropy of the universe 
should be realized inside the bubble.
This requirement can be satisfied by assuming, 
for example, a scalar field with a potential such as in the 
new inflation scenario but with a high potential barrier before 
the slow rolling inflationary phase\cite{BGT,YST95}. In addition, 
the high potential barrier keeps the bubble collision rate
small, hence the homogeneity and isotropy of the one-bubble universe is
not disturbed by the other nucleated bubbles.
Then the second inflation inside the bubble inflates the universe
and explains the flatness $\Omega_0\sim1$.
Linde and Mezhlumian has proposed another (perhaps more natural)
model of open inflation by introducing two scalar 
fields\cite{Lindea,Lindeb}, but the
essential feature of scenario is the same as the former. 

The interesting problem is the origin of fluctuations in this open 
inflationary scenario. Following the usual picture that quantum
fluctuations of the scalar field generate the initial density 
perturbations, we need to investigate the quantum state of
the scalar field after bubble nucleation.
According to previous investigations, the open inflation scenario 
shows interesting varieties of the fluctuations.
The bubble nucleation process can excite fluctuations of the
scalar field and may increase the power of density fluctuations 
on scale of spatial curvature \cite{YTS95,HAMA}. Also it has been
shown that peculiar discrete modes of fluctuations on super-curvature 
scale may exist \cite{STY95} and contribute to cosmic microwave
background (CMB) anisotropies in an open universe \cite{YST95,YB}.
Very recently, generation of another type of super-curvature 
perturbations which originate from the bubble wall perturbations 
has been discussed \cite{HAMA,Garriga,Bellido}. 

In this paper, we develop a method to calculate these 
varieties of fluctuations in the open inflationary universe
and evaluate the power spectrum of the resulting CMB
anisotropies. 
In order to perform a detailed analysis of the power specturm, 
it is necessary to specify the model to some extent. 
Here we consider two classes of models. 
One is a single scalar field model (Model A) proposed by Bucher,
Goldhaber and Turok\cite{BGT} and by ourselves\cite{YST95}. 
The other is a two-field model (Model B) in which
the false vacuum decay and inflation inside the bubble are
governed by two different scalar fields, such as
the supernatural inflation proposed by Linde and
 Mezhlumian\cite{Lindeb}. 
For simplicity, we assume the difference in the vacuum energy density
between inside and outside the bubble is negligible in both models. 
Our method is based on the formalism we developed previously
for computing the mode functions during and after bubble nucleation
\cite{TSY94,TS94,STY95}.

This paper is organized as follows. In section 2 we describe our method
to calculate quantum fluctuations of the inflaton field.
We assume that the reaction to the geometry can 
be neglected, i.e., the nucleation 
occurs in the de Sitter spacetime background.
Then, in section 3, we investigate the evolution of fluctuations
inside the bubble in an open inflationary stage and 
derive the formulas to calculate 
the initial perturbation spectrum of the open universe
and the resulting CMB anisotropies.
In section 4 we evaluate the CMB anisotropies 
in two simple models and examine their viability. 
Section 5 is devoted to conclusion.
We adopts the units $c=\hbar=1$, and
the bar denotes the complex conjugate.

\section{Formalism}

In this section, we describe our formalism 
to investigate quantum field fluctuations inside a bubble.
Pioneering work on the quantum state of a nucleating bubble
was done by Rubakov \cite{Rubakov}, Kandrup \cite{Kandrup},
Vachaspati and Vilenkin \cite{VV}. Recently, we have 
developed a formalism to investigate the quantum state 
of a scalar field after false vacuum decay based on the WKB 
wave function of a multi-dimensional tunneling system \cite{TSY94,OUR}.
The formalism has been applied to the bubble nucleation 
which occurs on the Minkowski background, and the spectrum of field 
fluctuations after the decay has been studied \cite{YTS95,HAMA}.
The basic formalism has been extended to the case in which the bubble 
nucleation occurs on the de Sitter spacetime\cite{TS94}. 
There the effect of the non-trivial geometry of the instanton 
with gravity, i.e., the Coleman-De Luccia instanton\cite{DeLCol} 
was taken into account. 
At that time, however, the appropriate set of 
the mode functions to describe the initial vacuum state,
 which is expected to be the 
Bunch-Davies vacuum due to the sufficient inflation before the
tunneling, had not been known. 
Thus there was a technical difficulty in applying our formalism.
Recently, we have succeeded in describing the Bunch-Davies vacuum 
state in the spatially open chart\cite{STY95}. Thus
this problem has been overcome. 
Combining the results in these two papers\cite{TS94,STY95}, 
we now have a tool to hundle 
the quantum fluctuations after tunneling which includes the effect 
of the geometry of the Coleman-De Luccia instanton 
similar to the case in the Minkowski background\cite{YTS95}. 

In the present paper, we consider
simple open inflation models in which the geometry of the
Coleman-De Luccia instanton can be approximated by the 
pure de Sitter spacetime. 
That is we consider the case in which
the potential energy difference between
the two vacua of the tunneling filed is small.
We consider the action
\begin{equation}
 S=\int\biggl(-{1\over2}g^{\mu\nu}\partial_{\mu}\sigma 
 \partial_{\nu}\sigma-V(\sigma)\biggr)\sqrt{-g}d^4x
  +\int\biggl(-{1\over2}g^{\mu\nu}\partial_{\mu}\phi 
 \partial_{\nu}\phi-U(\sigma,\phi) 
 \biggr)\sqrt{-g}d^4x, 
\end{equation}
where the potential $V(\sigma)$ is assumed to have the form
as shown in Figs. 1(a), (b) to realize the false vacuum decay. 
As we mentioned above, we take account of the effect of gravity
only as a curved background which is assumed to be de Sitter space.
We denote the value of $\sigma$ at the false vacuum by $\sigma_F$.
The field $\phi$ is the inflaton in the nucleated bubble. 
We devide them into two parts as $\phi=\phi_B+\varphi$. 
$\phi_B$ is the seimi-classical, spatially homogeneous part 
of the field and $\varphi$ represents the quantum fluctuations 
around this background. 
We neglect the quantum fluctuations of $\sigma$ and denote
its $O(4)$-symmetric background by $\sigma_B$.
Due to the $O(4)$-symmetry of the instanton, 
$\sigma_B$ is a function of only one coordinate, say $\tau$.
In addition, we assume $\phi_B$ also respects the $O(4)$-symmetry.
Thus a model we have in mind is a two-field one,
with $\phi_B$ being constant at $\sigma_B=\sigma_F$.
Note that we can consider a single field model by
identifying $\varphi$ with the quantum fluctuations of $\sigma$
and letting $U(\sigma,\phi)=V''(\sigma)\phi^2/2$.

With these assumptions, the action for $\varphi$ reduces to
\begin{equation}
 S_{\varphi}=
\int\biggl(-{1\over2}g^{\mu\nu}\partial_{\mu}\varphi 
 \partial_{\nu}\varphi-{1\over 2}{\cal M}^2(\tau)\varphi^2 
 \biggr)\sqrt{-g}d^4x, 
\end{equation}
where 
\begin{equation}
 {\cal M}^2(\tau)={\partial^2 U\over \partial \phi^2}
        \left(\sigma_B(\tau),\phi_B(\tau)\right).
\end{equation}
As noted above, in the case of a single field model,
 $\phi_B$ and ${\cal M}^2(\tau)$ in the following 
discussion should be replaced by $\sigma_B$ and 
$\displaystyle 
 {\partial^2 V \over \partial \sigma^2}
        \left(\sigma_B(\tau)\right),
$
respectively.

Our formalism is summarized in the following. 
It is based on the WKB wave function which 
describes the tunneling decay in a multi-dimensional system 
\cite{TSY94,TS94}. 
The construction of the wave function was originally developed by 
Gervais and Sakita \cite{GS}.
The wave functional is written in terms of the bounce solution
$\sigma_B$,
which gives the semi-classical picture of the nucleation and 
expanding bubble, and a set of mode functions 
which describe the quantum fluctuations $\varphi$ on the bounce
solution\cite{TSY94}. The boundary condition for these mode functions
is determined from the fact that the wave function describes the false
vacuum state at $\sigma_B=\sigma_F$.
It then turned out that the procedure to obtain the appropriate set of
mode functions is equivalent to finding out a complete set of
them which are regular in one hemisphere of the Euclideanized de
Sitter space\cite{TS94,STY95}.
That is, to find a complete set of mode functions $v_{\bf k}$
 which obey
\begin{equation}
  \Bigl[\nabla^\mu\nabla_\mu -{\cal M}^2\Bigr]
v_{\bf k}(t,{\bf x})=0,
\end{equation}
in the Lorentzian region, and which are regular on the 
${\rm Im}\,t<0$ hemisphere.
Here, it is important to note that ``a complete set'' means 
a set of all modes which are properly
normalized by the Klein-Gordon norms on a Cauchy 
surface $\Sigma$ of spacetime,
\begin{eqnarray}
  \left\langle v_{{\bf k}}\,,v_{{\bf k}'}\right\rangle
:=&&-i \int_{\Sigma}d\Sigma_\mu g^{\mu\nu} 
  \left\{v_{\bf k} \partial_\nu {\bar v_{{\bf k}'}}
  -(\partial_\nu v_{{\bf k}})\bar v_{{\bf k}'}\right\}
\nonumber\\
  =&&\delta_{{\bf k}{\bf k}'}.
\end{eqnarray}
Once they are obtained,
the quantum fluctuations of the field are described
by the ``vacuum state", $|\Psi\rangle$, such that 
$\hat a_{\bf k}|\Psi\rangle=0$
for any ${\bf k}$ where the fluctuating field is
expressed as
\begin{equation}
  \hat\varphi_H=\sum_{\bf k} (v_{\bf k} \hat a_{\bf k} + 
 {\bar v_{\bf k}}\hat a_{\bf k}^\dagger),
\end{equation}
in the Heisenberg representation, with
$\hat a_{\bf k}$ and $\hat a_{\bf k}^\dagger$ being the
annihilation and creation operators, respectively.
Thus the mode functions $v_{\bf k}$ play the role of
the positive frequency functions.

To write down the equation for the mode functions,
we introduce the coordinates in the de Sitter spacetime
following Ref.\cite{STY95}.
The four-dimensional Euclidean de Sitter space is a four-sphere.
The metric is represented as 
\begin{equation}
ds_E^2=H^{-2}d\tau^2
 +a_E^2(\tau)(d\rho^2+\sin^2\rho ~d\Omega^2)],
\end{equation}
where $-\pi/2\leq\tau\leq\pi/2$, $0\leq\rho\leq\pi$
and $a_E(\tau)=H^{-1}\cos\tau$.
The mode functions are required to be regular on the
$0\leq\rho\leq\pi/2$ hemisphere.

The coordinate systems in the Lorentzian de Sitter space are 
obtained by analytic continuation as
\begin{eqnarray}
   &t_R=i(\tau-\pi/2) \vspace{1cm}& (t_R\geq0),
\nonumber\\
   &r_R=i\rho \vspace{1cm}& (r_R\geq0),
\nonumber\\
   &t_C=\tau \vspace{1cm}& (\pi/2\geq t_C\geq-\pi/2),
\nonumber\\
   &r_C=i(\rho-\pi/2) \vspace{1cm}& (\infty>r_C>-\infty),
\nonumber\\
   &t_L=i(-\tau-\pi/2)~  \vspace{1cm}&  (t_L\geq0),
\nonumber\\
   &r_L=i\rho \vspace{1cm}& (r_L\geq0).
\label{RCLco}
\end{eqnarray}
we find that each set of these coordinates covers three distinct parts of
the Lorentzian de Sitter spacetime, which we call the regions
$R$, $C$, and $L$ (see Fig.2).
The metrics in these three regions are given by 
\begin{eqnarray}
  &&ds^2_R=-H^{-2}dt_R^2+a^2(t_R)(dr_R^2+\sinh^2r_R\,d\Omega^2),
\nonumber\\
  &&ds^2_C=H^{-2}dt_C^2+a_E^2(t_C)(-dr_C^2+\cosh^2r_C\,d\Omega^2),
\nonumber\\
  &&ds^2_L=-H^{-2}dt_L^2+a^2(t_L)(dr_L^2+\sinh^2r_L\,d\Omega^2),
\label{dsmetric}
\end{eqnarray}
respectively, where $a(t)=H^{-1}\sinh t$.
We assign the region L to be in the false vacuum sea
and the region R to describe the open universe inside the bubble.

The equation for the bounce in the Euclidean region is given by
\begin{equation}
 {H^2\over a_E^3}{d\over d\tau}
  \left(a_E^3{d \sigma_B\over d\tau}\right)
  -{V'(\sigma_B)}=0.
\label{bounceeq}
\end{equation}
The equation in the Lorentzian region is given by the analytic
continuation of the coordinates as specified in Eqs.~(\ref{RCLco}).
The fluctuation field $\varphi$ obeys, in the Euclidean region,
\begin{equation}
  \biggl[{1\over a_E^3}{\partial\over\partial(H^{-1}\tau)}a_E^3 
  {\partial \over \partial (H^{-1}\tau)}
  -{1\over a_E^2}{\bf L}^2 -{{\cal M}^2(\tau)}
  \biggr]\varphi=0,
\label{eqinC}
\end{equation}
where 
\begin{equation}
  {\bf L}^2=-{1\over\sin^2\rho}{\partial\over\partial\rho}\biggl(
  \sin^2\rho{\partial\over\partial\rho}\biggr)
  -{{\bf L}^2_{\Omega}\over\sin^2\rho},
\end{equation}
and ${\bf L}^2_{\Omega}$ is the Laplacian on a unit two sphere. 

Setting $\varphi=a_E(\tau)^{-1}\chi_p(\tau)f_{pl}(\rho)Y_{lm}(\Omega)$,
the equations for $\chi_p$ and $f_{pl}$ become, 
respectively,
\begin{equation}
\left[(1-x^2){d^2\over dx^2}-2x{d\over dx}+{p^2\over1-x^2}+
2-{{\cal M}^2\over H^2}\right]\chi_p=0,
\label{Eqchi}
\end{equation}
where $x=\sin\tau$, and
\begin{equation}
 \left[-{1\over\sin^2\rho}{\partial\over\partial\rho}\biggl(
  \sin^2\rho{\partial\over\partial\rho}\biggr)
  +{l(l+1)\over\sin^2\rho}+(p^2+1)\right]f_{pl}=0.
\label{Eqfpl}
\end{equation}
The requirement that the mode functions be regular at $\rho=r_L=r_R=0$
fixes the form of $f_{pl}$ to be\cite{STY95}
\begin{eqnarray}
f_{pl}(r)
&=&{\Gamma(ip+l+1)\over\Gamma(ip+1)}{p\over\sqrt{\sinh r}}
\,P_{ip-1/2}^{-l-1/2}(\cosh r)
\nonumber\\
&=&(-1)^l\sqrt{2\over\pi}{\Gamma(-ip+1)\over\Gamma(-ip+l+1)}
\sinh^lr{d^l\over d(\cosh r)^l}\left({\sin pr\over\sinh r}\right),
\label{fpldef}
\end{eqnarray}
where $r=i\rho$ in the Euclidean region.
The gamma function factor in front is attached so that
the functions $Y_{plm}(r,\Omega):=f_{pl}(r)Y_{lm}(\Omega)$ 
become properly normalized harmonics on the hyperbolic slices
in the region R or L:
\begin{equation}
\int_0^\infty dr\sinh^2r\int d\Omega\,
Y_{plm}(r,\Omega)\overline{Y_{p'l'm'}(r,\Omega)}
=\delta(p-p')\delta_{ll'}\delta_{mm'}\,.
\end{equation}

As we have noted before, a complete set of mode functions should be
properly normalized on a Cauchy surface. Therefore we consider
the mode functions in the region C, since it is the region in which
complete Cauchy surfaces exist. It should be noted that 
$f_{pl}$ play the role of the positive frequency functions there.
Then the remaining tasks are (1) to find the possible
spectrum of $p$ that may give finite Klein-Gordon norms and 
(2) to properly normalize the mode functions if they are
normalizable.

To accomplish the first task, we now consider the equation
for $\chi_p$, (\ref{Eqchi}). In the region C, the equation for
$\chi_p$ remains the same because $t_C=\tau$.
Introducing the conformal coordinate $\xi$ 
by the relation $x=\sin\tau=\tanh\xi$, 
the metric in the region C becomes
\begin{equation}
 ds^2=a_E^2(\xi)\left(-dr_C^2+d\xi^2+\cosh^2 r_C~d\Omega^2\right), 
\label{conform}
\end{equation}
where $a_E=(H\cosh\xi)^{-1}$ and Eq.~(\ref{Eqchi}) is rewritten as
\begin{equation}
  \biggl[-{d^2 \over d\xi^2}+U(\xi) \biggr]\chi_p
=p^2\chi_p\,,
\label{eqinCe}
\end{equation}
where 
\begin{equation}
  U(\xi)={{\cal M}^2/H^2-2\over \cosh^2\xi}\,.
\label{Uxi}
\end{equation}
This resembles the Schr\"odinger equation with the 
potential $U(\xi)$ and the energy $E=p^2$.
Thus finding possible values of $p$ is equivalent to
solving the eigenvalue problem of the above equation.
Since $U(\xi)\to0$ at $\xi\to\pm\infty$,
the spectrum of modes is continuous for $p^2\geq0$.
On the other hand, if ${\cal M}^2 <2 H^2$ for some 
interval of $\xi$, the potential has a valley and
some discrete modes which correspond to bound
 states may appear for $p^2=-\Lambda^2<0$.
In fact, it has been shown that there exists a discrete eigenvalue 
of $\Lambda^2$ in two different simple cases.
One is a mode which appears when we consider a scalar field with a 
constant small mass $(<\sqrt{2H^2})$ on the de Sitter 
spacetime\cite{STY95}. Its origin is intrinsic to 
the spacetime structure of the de Sitter universe.
As long as the volume with ${\cal M}^2<2 H^2$ is large enough, 
an analogous mode is expected to exist in general.
Let us call it the de Sitter super-curvature 
mode.\footnote{The spatial
 harmonics behave as $Y_{plm}(r)\propto e^{-r}$ for $p^2>0$ . 
As $r=1$ corresponds to the scale of spatial curvature, 
fluctuations described by the harmonics with $p^2>0$ 
represents those which
decay exponentially on scales larger than the curvature 
scale. In contrast, harmonics with $p^2=-\Lambda^2<0$
behave as $Y_{\Lambda lm}\propto 
e^{(\Lambda-1)r}$, hence describe fluctuations over scales
larger than the curvature scale\cite{LW,GLLW}. 
The name ``super-curvature" originates from this fact.}
The other is the mode that appears when we cosider the fluctuations 
of the tunneling field itself\cite{HAMA}. 
It describes the fluctuation of the wall.
Since it turns out that the analysis of the wall fluctuation mode 
can be simplified compared with the other discrete modes, 
below we consider the continuous modes, the de Sitter super-curvature 
mode, and the wall fluctuation mode separately in sequence.
{}For simlicity, 
we assume that $\cal M$ is constant in the regions R and L 
and set
\begin{equation}
{\cal M}^2=\cases{m^2~&in R\,,\cr M^2~&in L\,,} 
\end{equation}
in the following discussion. 

\subsection{Continuous modes}

We first consider the continuous spectrum mode functions 
of $p^2>0$. For these modes, it has been shown that the Klein-Gordon
norms can be evaluated by the sum of those on the hyperbolic slices on
the regions R and L\cite{STY95}. 
Since in the regions R and L, ${\cal M}^2$ is supposed to be 
constant, 
$\chi_p$ is given by
an associated Legendre function $P^{ip}_{\nu-1/2}(z)$ 
or $P^{-ip}_{\nu-1/2}(z)$ with $\nu=\sqrt{9/4-{\cal M}^2/H^2}$. 
To construct the normalized mode functions,
we begin with the following mode functions in the regions R and L:  
\begin{eqnarray}
&&u_{plm}^{(R)}={1\over a(t_R)}\,\chi_{p}^{(R)}(t_R)Y_{plm}(r_R,\Omega),
\nonumber \\
&&u_{plm}^{(L)}={1\over a(t_L)}\,\chi_{p}^{(L)}(t_L)Y_{plm}(r_L,\Omega),
\end{eqnarray}
where
\begin{eqnarray}
\chi_{p}^{(R)}(t_R) & = & P^{ip}_{\nu'}(z_R);\quad z_R=\cosh t_R\,,
\nonumber \\
\chi_{p}^{(L)}(t_L) & = & P^{ip}_{\mu'}(z_L);\quad z_L=\cosh t_L\,,
\end{eqnarray}
with $\nu'=\sqrt{9/4-m^2/H^2}-1/2$ and $\mu'=\sqrt{9/4-M^2/H^2}-1/2$.
The norms of these functions are give by
\begin{eqnarray}
[(\chi_{p}^{(R)},\chi_{p}^{(R)})]^{(R)} & := &
i(z_R^2-1)\left\{{d\chi_{p}^{(R)}\over dz_R}\bar\chi_{p}^{(R)}
 -\chi_{p}^{(R)}{d\bar\chi_{p}^{(R)}\over dz_R}\right\}
={2\over \pi}\sinh \pi p, 
\cr 
[(\chi_{p}^{(L)},\chi_{p}^{(L)})]^{(L)} & := &
i(z_L^2-1)\left\{{d\chi_{p}^{(L)}\over dz_L}\bar\chi_{p}^{(L)}
 -\chi_{p}^{(L)}{d\bar\chi_{p}^{(L)}\over dz_L}\right\}
={2\over \pi}\sinh \pi p.  
\label{chinorm}
\end{eqnarray}
If we analytically continue $\chi_p^{(R)}$ to the region L by
solving Eq.~(\ref{eqinCe}) in the region C, $\chi_p^{(R)}$ 
will be expressed in terms of a linear 
combination of $\chi_p^{(L)}$ and $\chi_{-p}^{(L)}$.
Hence we consider the mode function,
\begin{equation}
\chi_{p}=\chi_{p}^{(R)}=\alpha_p \chi_{-p}^{(L)}+\beta_p \chi_{p}^{(L)}.
\end{equation}

{}From the property of the analytic continuation\cite{STY95} and the 
reality of the Eq.~(\ref{eqinCe}), 
there exists a symmetry,
\begin{eqnarray}
 \bar \chi_p^{(R)}  & = & e^{-\pi p}\chi_{-p}^{(R)},
\nonumber \\
 \bar \chi_p^{(L)}  & = & e^{-\pi p}\chi_{-p}^{(L)}, 
\end{eqnarray}
in the region C. 
Then we obtain
\begin{equation}
 \bar \alpha_p = e^{-2\pi p} \alpha_{-p},\quad 
 \bar \beta_p = \beta_{-p}. 
\end{equation}
Further, defining the Wronskian by 
\begin{equation}
 ((u,v)):=u {dv \over d\xi}-{du \over d\xi}v, 
\end{equation}
$\alpha_p$ and $\beta_p$ are expressed as
\begin{equation}
\alpha_p={((\chi_p^{(L)},\chi_{p}^{(R)}))
\over((\chi_p^{(L)},\chi_{-p}^{(L)}))}\,,
\quad
\beta_p={((\chi_p^{(R)},\chi_{-p}^{(L)}))
\over((\chi_p^{(L)},\chi_{-p}^{(L)}))}\,,
\label{alphbeta}
\end{equation}
and we have
\begin{equation}
((\chi_p,\bar\chi_p))=((\chi_p^{(R)},\bar\chi_p^{(R)}))
=\left(e^{2\pi p}|\alpha_p|^2-|\beta_p|^2\right)
((\bar\chi_{p}^{(L)},\chi_p^{(L)})).
\end{equation}
Also from the analytic continuation 
of the Eq.~(\ref{chinorm}) we obtain
\begin{equation}
((\chi_p^{(R)},\bar\chi_p^{(R)}))
=((\bar\chi_p^{(L)},\chi_p^{(L)})). 
\end{equation}
Thus we find
\begin{equation}
 X:=\vert\beta_p\vert^2=e^{2\pi p}\vert\alpha_p\vert^2-1.
\end{equation}
Now define
\begin{equation}
[(\chi_1,\chi_2)]:=[(\chi_1,\chi_2)]^{(R)}+
                   [(\chi_1,\chi_2)]^{(L)},
\end{equation}
which gives the contribution of the function $\chi_p$ to the total
Klein-Gordon norm.
Then we have
\begin{eqnarray}
K_{++} &:= & [(\chi_p,\chi_p)]
 = {4\over\pi}e^{-\pi p}\sinh^2 \pi p\,(1+X),
\nonumber \\
K_{+-} &:= & [(\chi_{p},\chi_{-p})]
={4\over\pi}e^{\pi p}\sinh^2\pi p\,\alpha_p\beta_p\,,
\nonumber \\
K_{-+} &:= & [(\chi_{-p},\chi_{p})]
={4\over\pi}e^{\pi p}\sinh^2\pi p\,\bar\alpha_p\bar\beta_p\,,
\nonumber \\
K_{--} &:= & [(\chi_{-p},\chi_{-p})]
 = {4\over\pi}e^{\pi p}\sinh^2 \pi p\,(1+X),
\end{eqnarray}
and
\begin{equation}
|K_{+-}|^2=\left({4\over\pi}\sinh^2 \pi p\right)^2X(1+X).
\end{equation}

To find the orthonormalized mode functions, we must find a
matrix $M_{\sigma\sigma'}$ which diagonalizes 
the matrix $K_{\sigma\sigma'}$ 
($\sigma,\sigma'=+,-$).\footnote{Instead of allowing $p$
to have the range $-\infty<p<\infty$, we restrict it in the 
range $0<p<\infty$ but doubling the degrees of freedom with
$\sigma=\pm1$.}
It can be chosen as
\begin{equation}
  {M}_{\sigma\sigma'}=
  \left(\begin{array}{cc}
 \displaystyle {-K_{-+}\over\sqrt{|K_{+-}|^2+(K_{++}-\lambda_{+})^2}}
 & 
 \displaystyle
 {K_{++}-\lambda_{+}\over\sqrt{|K_{+-}|^2+(K_{++}-\lambda_{+})^2}} 
\\
 \displaystyle {-K_{-+}\over\sqrt{|K_{+-}|^2+(K_{++}-\lambda_{-})^2}}
 & 
 \displaystyle
 {K_{++}-\lambda_{-}\over\sqrt{|K_{+-}|^2+(K_{++}-\lambda_{-})^2}}
  \end{array}\right),
\end{equation}
where $\lambda_{\pm}$ are the eigenvalues of $K_{\sigma\sigma'}$,
\begin{equation}
  \lambda_{\pm}
={1\over2}\left((K_{++}+K_{--})
  \pm\sqrt{(K_{++}-K_{--})^2+4|K_{+-}|^2}\,\right).
\end{equation}
Note that the matrix $K_{\sigma\sigma'}$ is hermitian with
positive eigenvalues, as it should be.
Then the orthonormalized mode functions are given by
\begin{eqnarray}
v_{p\sigma lm}&=&V_{p\sigma}Y_{plm}\,;
\nonumber\\
 V_{p\sigma}&=&{1\over a(t)\sqrt{\lambda_\sigma}}
         \left(M_{\sigma+}\,\chi_{p}+M_{\sigma-}\,\chi_{-p}\right).
\label{contmode}
\end{eqnarray}

\subsection{de Sitter super-curvature mode}

Next we consider the discrete mode that arises due to
the spacetime structure of de Sitter space. 
As mentioned above we consider the wall perturbation mode seperately later. 
For convenience, we set $p=i\Lambda$ and look for a possible value of 
$\Lambda$.
When there is no bubble wall, it has been shown that
there arises a mode with $\Lambda=\nu'$ ($\nu'=\sqrt{9/4-m^2/H^2}-1/2$)
when $\nu'>0$ ($m^2<2H^2$)\cite{STY95}. 
In the present case, because the wall is present and the mass changes 
from $M$ to $m$, 
such a mode may or may not exist, depending on the function ${\cal M}^2$ 
even if $m^2<2H^2$. 

In the region C, we consider the solution of the form,
\begin{equation}
  u_{\Lambda lm}(r_C,t_C,\Omega)={1\over a_E(t_C)}\,
  \chi_{\Lambda}(t_C)\,
  {{P}^{-l-1/2}_{-\Lambda-1/2}(i\sinh r_C)\over\sqrt{i\cosh r_C}}
  \, Y_{lm}(\Omega),
\label{Vmode}
\end{equation}
where we set the form of $\chi_\Lambda$ as
\begin{equation}
\chi_{\Lambda}
  =\cases{
  \displaystyle
   \alpha_\Lambda P_{\mu'}^{\Lambda}(\sin t_C-i0)
  +\beta_\Lambda P_{\mu'}^{-\Lambda}(\sin t_C-i0),
& ($t_C\rightarrow -\pi/2$),
\cr
  \displaystyle
  P_{\nu'}^{-\Lambda}(\sin t_C-i0)
 +\gamma_\Lambda P_{\nu'}^{\Lambda}(\sin t_C-i0),
& ($t_C\rightarrow \pi/2$), \cr
}
\end{equation}
near the both boundaries of the region C. 
The regularity condition at $t_C=-\pi/2$ is 
that the mode function should be less singular
than $(t_C+\pi/2)^{-1}$, hence $\chi_\Lambda\to0$.
Using the asymptotic behavior of $P^\mu_\nu(z)$ at $z\to-1$
\cite{Magnus},
this requires the ratio of $\beta_\Lambda$ to 
$\alpha_\Lambda$ to be
\begin{equation}
\beta_\Lambda={\sin\pi\mu'\over\pi}\Gamma(1+\Lambda+\mu')
\Gamma(\Lambda-\mu')\,\alpha_\Lambda\,.
\label{regcond}
\end{equation}
Similarly, the regularity condition at $t_C= \pi/2$ 
demands that $\gamma_\Lambda$ should vanish:
\begin{equation}
\gamma_\Lambda=0.
\end{equation}
Then we need to solve the Eq.~(\ref{eqinCe}) with the above
boundary conditions in both ends.
Thus the problem is to find the eigenvalue 
$\Lambda$ and the corresponding normalized eigenmode. 
This eigenvalue problem will be solved for a simple model in
section 4.

Now let us consider the normalization of the super-curvature mode.
The Klein-Gordon inner products of the mode functions
$u_{\Lambda lm}$ are calculated in the region C as
\begin{eqnarray}
  &&\left\langle u_{\Lambda lm}(r_C,t_C,\Omega),
  u_{\Lambda l'm'}(r_C,t_C,\Omega)\right\rangle
\nonumber\\
&&\vspace{1cm}={-i\cosh^2r_C}
\int_{-\infty}^{\infty}d\xi\, a_E^2 
  \int d\Omega\left\{u_{\Lambda lm}\,
 {\partial\,\overline{u_{\Lambda l'm'}}\over\partial r_C}\,
-{\partial\,u_{\Lambda lm}\over\partial r_C}
\overline{u_{\Lambda l'm'}}\right\}\,
\nonumber\\
&&\vspace{1cm}=
  {2 N_{\Lambda}\over\Gamma(\Lambda+l+1)\Gamma(-\Lambda+l+1)}
  \,\delta_{ll'}\delta_{mm'}\,,
\end{eqnarray}
where
\begin{equation}
  N_{\Lambda}:= \int_{-\infty}^{\infty}
{d\xi}\left|\chi_{\Lambda}\right|^2.
\end{equation}
Thus, the normalized mode functions $v_{\Lambda lm}$
are given by
\begin{equation}
v_{\Lambda lm}
=\sqrt{\Gamma(\Lambda+l+1)\Gamma(-\Lambda+l+1)\over2N_\Lambda}\,
u_{\Lambda lm}\,.
\end{equation}
In particular, by analytic continuation to the 
region R (inside the bubble), we obtain the de Sitter
super-curvature mode functions in the open universe as
\begin{equation}
  v_{\Lambda lm}(t_R,r_R,\Omega)={1\over\sqrt{N_\Lambda}}\,
  {P^{-\Lambda}_{1}(\cosh t_R)\over H^{-1}\sinh t_R}\,
 {\cal Y}_{\Lambda lm}(r_R,\Omega)\,,
\label{desmode}
\end{equation}
where
\begin{equation}
  {\cal Y}_{\Lambda lm}:=
\sqrt{\Gamma(\Lambda+l+1)\Gamma(-\Lambda+l+1)\over2}
{P^{-l-1/2}_{\Lambda-1/2}(\cosh r_R)\over\sqrt{\sinh r_R}}Y_{lm}(\Omega).
\label{calYdef}
\end{equation}

\subsection{Wall fluctuation mode}

Recently, several authors have discussed the effect of
the bubble wall fluctuations\cite{HAMA,Garriga,Bellido}.
The wall fluctuation mode appears when we consider 
the quantum flucuations of the tunneling field itself, 
which is the case of a single field model.
As mentioned before, 
${\cal M}^2$ should be regarded as $V''(\sigma_B)$ in this case. 

Since the wall fluctuation mode is one of the discrete modes, 
the formula obtained in the previous subsection is applicable. 
However, besides the fact that the origin 
of the discrete spectrum is different from the 
de Sitter super-curvature mode, 
the corresponding solution of the Eq.~(\ref{eqinCe}), $\chi_W$,
can be written down formally in this special case.
Hence we consider this mode separtely in this subsection. 

It is well known that the time derivative of the bounce 
solution satisfies the equation of fluctuations,
related to the zero mode problem\cite{CC}.
In fact, we can derive the following equation directly 
from Eq.~(\ref{bounceeq}),
\begin{equation}
  \biggl[{H^2\over a_E^3}{d\over dt_C}
     a_E^3 {d\over dt_C}
  -{3\over a_E^2} -{V''(\sigma_B)}
  \biggr] {d\sigma_B\over dt_C}=0.
\label{eqinCdd}
\end{equation}
This is just the equation of fluctuations, i.e., eq.(\ref{eqinC})
with the eigenvalue of $-{\bf L}^2$ given by $1+p^2=-3$,
or $\Lambda^2=-p^2=4$. 
Then if they have finite Klein-Gordon norms, they will
contribute to the quantum fluctuations inside the bubble.

Therefore let us consider the mode functions
\begin{equation}
  u_{W,lm}={d\sigma_B\over d(H^{-1}t_C)}
  {P_{3/2}^{-l-1/2}(i\sinh r_C)\over\sqrt{i\cosh r_C}}
  \,Y_l^m(\Omega).
\end{equation}
Calculating the Klein-Gordon norms, we obtain
\begin{equation}
\left\langle u_{W,lm}\,,u_{W,l'm'}\right\rangle
={2{\cal N}_W\over\Gamma(l+3)\Gamma(l-1)}\,\delta_{ll'}\delta_{mm'}\,,
\end{equation}
where
\begin{equation}
{\cal N}_W:=H\int_{-\pi/2}^{+\pi/2} dt_C\, a_E\,
   \left\vert {d\sigma_B\over dt_C}\right\vert^2=
  \int_{-\infty}^{+\infty} d\xi
  \left\vert {d\sigma_B\over d\xi}\right\vert^2.
\end{equation}
Note that ${\cal N}_W$ has the dimension of $(\hbox{mass})^2$.
As one can readily see, the Klein-Gordon norms vanish for
$l=0$ and $1$. One may think that they give a divergent 
contribution to the fluctuations. 
However, they simply represent the temporal and spatial translations
of the origin of the bounce solution, i.e., they are the zero 
modes. Hence we do not have to take these modes into account 
\cite{CC}. On the other hand, the modes with $l\geq2$
do have finite Klein-Gordon norms and they contribute
to the fluctuations\cite{HAMA}.
We note that in the exact thin-wall limit, 
the amplitude of these modes are non-zero only on the bubble wall
and they do not contribute to the fluctuations
in the open universe (region R).
It is crucially important that $\dot\sigma_B$ is not exactly zero
off the wall, however small it may be.
Then the normalized mode functions are given by
\begin{equation}
v_{W,lm}=\sqrt{\Gamma(l+3)\Gamma(l-1)\over2{\cal N}_W}\,
u_{W,lm}\,.
\end{equation}
Analytically continuing the mode functions into the region R,
we have in the open universe,
\begin{equation}
  v_{W,lm}(r_R,t_R,\Omega)=
  {1\over \sqrt{{\cal N}_W}}{d\sigma_B\over d(H^{-1}t_R)}
  {\cal Y}_{\Lambda=2,lm}(r_R,\Omega),
\label{wallmode}
\end{equation}
where ${\cal Y}_{\Lambda lm}$ has been defined in Eq.~(\ref{calYdef}).
An important property of this mode is that the second rank tensor constructed 
from the spatial harmonic function happens
to be transverse-traceless \cite{HAMA}. Furthermore, the gauge-invariant
density perturbation $\Delta$, which represents the density perturbation
on the comoving hypersurface, vanishes identically.
Thus it may be viewed as a kind of
gravitational wave mode\cite{Garriga}.
 However, in order to treat all the modes
on an equal footing, we do not take that view here. 
\vspace{5mm}

To summarize this section, we have discussed the three distinct
kinds of mode functions which describe 
quantum fluctuations in open inflation models.
The first one is the usual modes having continuous spectrum. 
The other two are modes with discrete spectra;
the de Sitter super-curvature mode and the wall fluctuation mode.
These quantum fluctuations give rise to cosmological metric
perturbations which will be reflected in observational quantities.
We will discuss this aspect of the quantum fluctuations
in the next section.

\section{Initial fluctuations in open universe and
large angle CMB anisotropies}

Once we have the mode functions which describe the
quantum state of the inflaton field inside the bubble,
we can investigate the evolution of resulting cosmological 
perturbations in the open inflationary stage and 
the subsequent stage of the open Friedmann universe, 
which can be compared with observational data. 
In this section, we first consider the initial condition of the 
cosmological perturbations generated during the
open inflationary stage, based on the results obtained in the
previous section.
Since the inflaton is almost massless during the open
inflationary stage inside the bubble, we set $m^2=0$.
Then we relate it to the spectrum of 
large angular scale CMB anisotropies.
Below, we work in the region R which describes the open universe
inside the bubble. We write the line element as 
\begin{eqnarray}
  ds^2&=&-dt^2+a^2(t)\gamma_{ij}dx^idx^j
\nonumber\\
&=&a^2(\eta) \Bigl[-d\eta^2+\gamma_{ij}dx^idx^j\Bigr],
\label{linee}
\end{eqnarray}
where $a$ is the scale factor, $t$ is the cosmological proper 
time,\footnote{Up to now we have used the non-dimensional time 
normalized by the Hubble parameter. But from now on,
we recover the dimension of time in $t$.}
$\eta$ is the conformal time, and
\begin{equation}
  \gamma_{ij}dx^idx^j=dr^2+\sinh^2r d\Omega^2.
\end{equation}
At the open inflationary stage, the scale factor is 
assumed to be given by
\begin{equation}
a={\sinh Ht\over H}={1\over H\sinh(-\eta)}\,,
\end{equation}
where $H$ is sufficiently slowly varying in time.

Just after the bubble is formed, or in the spacetime region close to
the boundary light cone of the region R, the universe is dominated by
the curvature term, or the matter energy density can be
neglected. Hence the metric perturbations induced by
the quantum fluctuations of the scalar field are negligible.
On the other hand,
there exists a time slicing on which the effect of
metric perturbations is minimal, called the flat hypersurfaces.
This implies that the fluctuations described by
the mode functions we have obtained in the previous section
may be regarded as those on the flat hypersurfaces.
It is easy to show that the curvature perturbation
on the comoving hypersurface ${\cal R}_c$ is related to the scalar field 
perturbation $\varphi$ on the flat hypersurface as
${\cal R}_c=-\dot a/(a\dot\phi_B)\varphi$ \cite{LS}.
Assuming the perturbation is adiabatic, ${\cal R}_c$ completely
determines the metric and matter perturbations of scalar type
(i.e., those which can be expanded in terms of scalar harmonics on the
 3-space).
An important property of ${\cal R}_c$ is that it remains constant
on superhorizon scales as long as the perturbation is adiabatic,
irrespective of the background spatial curvature of the universe
(see Appendix A).
Hence what we need to evaluate is the amplitude of ${\cal R}_c$
for each mode when the mode passes the horizon scale.
Assuming $\dot\phi_B$ and $H$ are sufficiently slowly varying at
the inflationary stage, we may then evaluate ${\cal R}_c$ at the limit
$\eta\to0$ ($a\to\infty$). Thus we have
\begin{equation}
{\cal R}_c=-{H\over\dot\phi_B}\varphi\,;\quad \eta\to0\,.
\label{Rc}
\end{equation}

The spectrum of the curvature perturbation on the comoving 
hypersurfaces is given as follows.
Expanding the curvature perturbation by modes
\begin{eqnarray}
  {\cal R}_c&=&\sum_{\sigma lm} \int_0^\infty dp 
~{\cal R}_{p\sigma}Y_{plm}(r,\Omega)
\nonumber\\
      && +\sum_{lm} ~{\cal R}_\Lambda {\cal Y}_{\Lambda lm}(r,\Omega)
      +\sum_{lm} ~{\cal R}_W {\cal Y}_{2lm}(r,\Omega),
\label{calRexp}
\end{eqnarray}
the power spectrum of the curvature perturbation
is given by $\bigl|{\cal R}_p\bigr|^2
:=\sum_\sigma\bigl|{\cal R}_{p\sigma}\bigr|^2$ 
for continuous modes, by $\bigl|{\cal R}_\Lambda\bigr|^2$ for 
de Sitter super-curvature mode, and by $\bigl|{\cal R}_W\bigr|^2$ 
for wall fluctuation mode.

First consider the continuous modes. From Eqs.~(\ref{contmode})
and (\ref{Rc}), we find after some algebra,
\begin{equation}
\bigl|{\cal R}_{p}\bigr|^2
=\left({H^2\over\dot\phi_B}\right)^2{\coth\pi p\over2p(1+p^2)}
(1-Y),
\label{Rspec}
\end{equation}
where
\begin{equation}
Y={\Gamma(2-ip)\over\Gamma(2+ip)}\,
{e^{-\pi p}\beta_p\over2\cosh\pi p\,\bar\alpha_p}
+ \hbox{c.c.}\,.
\end{equation}
It may be worthwhile to compare this result with the spectra
of the Bunch-Davies vacuum state \cite{YST95,YB}
and the conformal vacuum state\cite{LS,RPb}. 
In these case we have 
\begin{eqnarray}
\bigl|{\cal R}_{p}\bigr|_{BD}^2
&=&\left({H^2\over\dot\phi_B}\right)^2{\coth\pi p\over2p(1+p^2)}\,,
\label{BDspec}
\\
\bigl|{\cal R}_{p}\bigr|_{C}^2
&=&\left({H^2\over\dot\phi_B}\right)^2{1\over2p(1+p^2)}\,,
\label{confspec}
\end{eqnarray}
respectively. The result of our model (\ref{Rspec}) differs
from that of the Bunch-Davies one in the last factor $(1-Y)$,
which represents the effect of the mass difference
outside the wall. However, as $Y\to0$ for $p\to\infty$,
the spectra in all the three cases behave 
as $\bigl|{\cal R}_p\bigr|^2\propto 1/p^3$ for $p\gg1$, 
which shows the perturbations have the 
Harrison-Zel'dovich spectrum on scales smaller than the curvature
scale, irrespective of the choice of vacuum.
 The difference appears on and above the curvature scale,
and it may become important when considering
large angular scale CMB anisotropies.

As for the de Sitter super-curvature mode,
the spectral amplitude is given from Eq.~(\ref{desmode}) as
\begin{equation}
\bigl|{\cal R}_\Lambda\bigr|^2
=\left({H^2\over\dot\phi_B}\right)^2
{1\over N_\Lambda\Gamma(2+\Lambda)^2}\,.
\label{Rdes}
\end{equation}
We note that, in the Bunch-Davies vacuum limit, we have
$\Lambda\to1$ and $N_\Lambda\to1/2$, which gives
\begin{equation}
\bigl|{\cal R}_\Lambda\bigr|_{BD}^2
=\left({H^2\over\dot\phi_B}\right)^2{1\over2}\,.
\end{equation}

The amplitude of the wall fluctuation mode is given from
Eq.~(\ref{wallmode}). We identify $\sigma_B$ with $\phi_B$
and note that $d\sigma_B/dt_R=H^{-1}\dot\sigma_B$.
Then
\begin{equation}
\bigl|{\cal R}_W\bigr|^2={H^2\over{\cal N}_W}\,.
\label{Rwall}
\end{equation}
One sees it is quite different from those of the other modes;
it contains no information of how the field evolves at the open
inflationary stage but is completely determined by the part of 
the potential which governs the false vacuum decay.

Now we consider the CMB anisotropies
predicted in our open inflation models using the initial conditions 
obtained in the above. 
The power spectra of CMB anisotropies
in open models have been well investigated in the case when
the scalar field is in the conformal vacuum \cite{LS,RPb,Kamion,Gorski}
or in the Bunch-Davies vacuum \cite{YST95,YB}.
Since we expect the difference between the present models and
the previously investigated models arises only on large angular 
scales, we focus on low mutipoles ($l<20$) of the spectrum.
Concerning the wall fluctuation mode, as we have mentioned, 
it has a very interesting property
that the perturbation in the spatial curvature induced by it is
transverse-traceless\cite{HAMA}.
Hence it may be regarded as a tensor mode perturbation.
Garriga have investigated the evolution of this mode in the
open inflationary universe from this point of view\cite{Garriga}. 
However, in this paper we treat it as a usual scalar perturbation
except for the fact that $p^2=-4$. For completeness, a proof that
both approaches are equivalent is given in Appendix B.

{}From the perturbed Einstein equations, the evolution equation of
${\cal R}_c$ becomes
\begin{equation}
{\cal R}_c''+2{a'\over a}{\cal R}_c'-K{\cal R}_c={\cal S},
\end{equation}
where the prime denotes the conformal time derivative,
$K=-1$ for an open universe and
the source term $\cal S$ can be neglected if
 the universe is matter-dominated (see Appendix A).
Since ${\cal R}_c=const.$ at the very early stage of the hot Friedmann
universe, the relevant solution of it is the so-called growing mode
solution.
Here we consider ${\cal R}_c$ on sufficiently large scales 
that it re-enters the horizon after the universe becomes
matter-dominated.
In this case, the line element is described by 
Eq.~(\ref{linee}) with the scale factor 
$a(\eta)\propto\cosh\eta-1 ~(\eta>0)$. 
Then we obtain
\begin{equation}
{\cal R}_c={\cal R}_{c,i}{3(\eta\sinh\eta-2\cosh\eta+2)\over(\cosh\eta-1)^2}\,,
\label{calRsol}
\end{equation}
where ${\cal R}_{c,i}$ is the initial amplitude determined at the
inflationary stage of the universe, the spectrum of which is
given by Eqs.~(\ref{Rspec}), (\ref{Rdes}) and (\ref{Rwall}).

Given the solution (\ref{calRsol}), we can 
compute the temperature fluctuations in the open universe 
in the gauge-invariant formalism \cite{GSS}.
For this purpose, we consider the gauge-invariant variables
$\Psi$ and $\Phi$ which describe the gravitational potential perturbation and
the curvature perturbation, respectively, on Newtonian slicing\cite{KS}.
We have $\Psi+\Phi=0$ in the present case. Then $\Phi$
 is related to ${\cal R}_c$ as (see Appendix A)
\begin{equation}
\Phi={\cal R}_c+{a'\over a}{\cal R}_c'\,,
\end{equation}
which gives
\begin{equation}
  \Phi(\eta)={\cal R}_{c,i}F(\eta),
\end{equation}
where
\begin{equation}
  F(\eta)={3(\sinh^2\eta-3\eta\sinh\eta+4\cosh\eta-4)\over(\cosh\eta-1)^3}.
\end{equation}
Note that we have $\Phi={3\over 5}{\cal R}_c$ in the limit $\eta\to0$.
The potential perturbation gives rise to CMB anisotropies
on large angular scales. Writing the temperature autocorrelation 
function in the form,
\begin{equation}
  C(\theta)={1\over4\pi}\sum_l(2l+1)C_l {\rm P}_l(\cos\theta),
\end{equation}
the multipole moment $C_l$ is given by a sum of contributions from 
the continuous modes and the two super-curvature modes, 
\begin{equation}
  C_{l}=C_{l}^{\rm (C)}+C_{l}^{\rm (S)}+C_l^{\rm (W)}.
\end{equation}
Assuming that the conformal time of the last scattering surface
is sufficiently small, $\eta_{LSS}\ll1$, but the universe is
well matter-dominated by that time, we obtain\cite{GSS}
\begin{eqnarray}
  C_{l}^{\rm (C)}&=&\int_{0}^\infty dp 
 \bigl|{\cal R}_p\bigr|^2
  \biggl[{1\over3}F(\eta_{LSS})f_{pl}(\eta_0-\eta_{LSS})
  +2\int_{\eta_{LSS}}^{\eta_0}{dF(\eta')\over d\eta'} 
  f_{pl}(\eta_0-\eta')d\eta'\biggr]^2,
\\
  C_{l}^{\rm (S)}&=&
  \bigl|{\cal R}_\Lambda\bigr|^2
  \biggl[{1\over3}F(\eta_{LSS}){\tilde f}_{\Lambda l}(\eta_0-\eta_{LSS})
  +2\int_{\eta_{LSS}}^{\eta_0}{dF(\eta')\over d\eta'} 
  {\tilde f}_{\Lambda l}(\eta_0-\eta')d\eta'\biggr]^2,
\\
  C_{l}^{\rm (W)}&=&
  \bigl|{\cal R}_W\bigr|^2
  \biggl[{1\over3}F(\eta_{LSS}){\tilde f}_{2l}(\eta_0-\eta_{LSS})
  +2\int_{\eta_{LSS}}^{\eta_0}{dF(\eta')\over d\eta'} 
  {\tilde f}_{2l}(\eta_0-\eta')d\eta'\biggr]^2,
\end{eqnarray}
where $\eta_0$ is the conformal time at present,
$f_{pl}$ is the function given in Eq.~(\ref{fpldef})
and
\begin{equation}
\tilde f_{\Lambda l}=
\sqrt{\Gamma(\Lambda+l+1)\Gamma(-\Lambda+l+1)\over2}
{P^{-l-1/2}_{\Lambda-1/2}(\cosh r_R)\over\sqrt{\sinh r_R}}\,.
\end{equation}
Here a word of caution is appropriate. The redshift of the 
matter-radiation equal time is $z_{eq}\approx4.2\times10^4\Omega_0h^2$, 
where $h$ is the Hubble constant
normalized by $100\,{\rm km/sec/Mpc}$ and $h\alt1$ from observations. 
In addition, $\cosh\eta_{LSS}=1+(\Omega_0^{-1}-1)/(1+z_{LSS})$
where $z_{LSS}$ is the redshift of the
last scattering surface ($\sim1000$). Therefore
the above formula for $C_l$ becomes inaccurate for $\Omega_0\ll0.1$. 
We should also note that
we have put $\eta_{LSS}=0$ for simplicity in actual calculations.
Hence our results presented in the next section may have errors of
$\sim 10\%$ for $\Omega_0\sim0.1$.

Before closing this section, we mention that the formalism developed
in the previous two sections can be extended to more general cases. 
Namely, as long as the regions R
and L are described by de Sitter space, it is not necessary to have the
same Hubble parameter for these two regions. 
Let $H_R$ and $H_L$ denote the Hubble parameters in the region R and L,
respectively. Then it is easily recognized that all $H$'s 
appearing in the formulas in the present section should be identified
with $H_R$.
Further, in this case, 
as the scale factor is different from that of de Sitter 
space in the region C, but the formulas written in terms of the 
conformal coordinate $\xi$ do not contain $H$ explicitly 
except for Eq.~(\ref{eqinCe}), what we need to do is
to replace the potential in Eq.~(\ref{eqinCe}) by
\begin{equation}
 U(\xi)\to {1\over a_E}{d^2 a_E\over d\xi^2}-1 +a_E^2 {\cal M}^2,
\end{equation} 
where $\xi$ is now defined in such a way that the metric in the region C
takes the form (\ref{conform}) with $a_E(\xi)$ being a general function
of $\xi$.

\section{Spectrum of the CMB anisotropies predicted by 
some models}
In this section, we apply our formulas obtained in sections 2 and 3
to simple models of the open inflation scenario. 

\subsection{Single field model}

We first consider a model with a single scalar field 
with the potential as shown in Fig. 1(a). 
The tunneling field becomes the inflaton field after the tunneling. 
Thus the mass square of the 
fluctuation field $\varphi$ is given by
${\cal M}^2(\tau)=V''(\sigma_B)$. 
Its typical shape is rather complicated as shown in Fig.~3. 

Because of the complexity of ${\cal M}^2(\tau)$, 
the analysis of this model is difficult 
in general. 
However, the mass of the inflaton in the 
false vacuum ${M}$ must be large 
compared with $H$ in order that the tunneling is 
not dominated by the Hawking-Moss instanton\cite{HawMos} 
as explained in Ref.~\cite{Lindeb}. 
Furthermore, as there are many subtleties\footnote{ 
For example, the formalism developed in \cite{TS94} 
which we used as a given result in the present paper 
is not applicable to the case in which the wall is 
spreading so broadly that we cannot approximate 
$\sigma_B(\tau)$ by $\sigma_F$ for $\tau<0$. 
When $M^2\sim H^2$, the wall necessarily becomes broad.} 
in models with $M^2 \sim H^2$, 
we leave it as a future problem. 
Here we assume $M^2\gg H^2$. 

Then the potential barrier between the two asymptotic regions
 at $\xi\rightarrow \pm \infty$ 
is so high that the coherence between 
regions R and L are exponentially suppressed 
as long as we consider the modes with $p^2\ll M^2/H^2$, 
which is the case of our present interest. 
So one of the two modes corresponding to the 
same $p$ can be set to vanish at $\xi\to -\infty$ 
and the other at $\xi\to +\infty$. 
Then automatically they are orthogonal to each other. 
Since the field equation (\ref{eqinCe}) 
is real in the region C, we can choose the mode 
which vanish in the region L to be
real in the region C.
Then analytically continueing it to the region R,
the normalized mode which does not vanish in 
the region R can be written as 
\begin{equation}
 V_p^{(R)}
   =\left\{\begin{array}{ll}
     \displaystyle {\sqrt{\pi}\over 2 a(t_R)\sinh \pi p} 
      \left(e^{\pi p/2+i\delta_p}
       (z_R-ip) \left(z_R+1\over z_R-1\right)^{ip}+
         e^{-\pi p/2-i\delta_p}
       (z_R+ip) \left(z_R+1\over z_R-1\right)^{-ip} \right),&
     \quad (\hbox{in R})\\
     0,&
     \quad (\hbox{in L})
    \end{array}\right. ,   
\label{largeM}
\end{equation}
where we set $m=0$ and $\delta_p$ is a real constant which depends
on the detail of the potential. 
Using this mode function, the amplitude of 
${\cal R}_p$ is evaluated as 
\begin{equation}
\bigl|{\cal R}_{p}\bigr|^2
=\left({H^2\over\dot\sigma_B}\right)^2
   {\cosh\pi p+\cos 2 \delta_p \over 2p(1+p^2)\sinh\pi p}. 
\label{Mspec}
\end{equation}
Comparing this result 
with the spectra for the Bunch-Davies vacuum (\ref{BDspec})
and the conformal vacuum (\ref{confspec}), we find that 
the difference between the present spectrum and 
$\bigl|{\cal R}_{p}\bigr|^2_{BD}$ 
is no more greater than the difference between 
$\bigl|{\cal R}_{p}\bigr|^2_{BD}$ and $\bigl|{\cal R}_{p}\bigr|^2_{C}$.
As was clarified in \cite{YST95}, 
the effect of the difference between the Bunch-Davies vacuum 
and the conformal vacuum on $C_{\ell}^{(C)}$ 
is always negligiblly small independent of $\Omega_0$. 
Thus as far as the continuous modes are concerned, we do not have to 
perform further calculations.

Next we consider the discrete mode. 
As mentioned above, the wall fluctuation mode is always present
when we consider the flucuations of the tunneling field itself. 
Then the question is whether there exists another discrete mode
like the de Sitter super-curvature mode.
The answer is no as long as $M^2\gg H^2$.
The reason why is explained in the Appendix C.
In the general case, we do not have an answer to 
this question,\footnote{We note an issue related to this problem.
One can show that
the non-existence of another discrete mode in the range
 $-4<p^2<0$ is a sufficient condition for the uniqueness 
of the negative eigenvalue mode in the one-loop order 
calculation of the tunneling rate in the path integral approach
\cite{nm1,nm2,nm3}. But the uniqueness has not been proved 
for the tunneling on the de Sitter background as far as we know.}
but we do not discuss the de Sitter super-curvature mode any further.
We focus on the wall fluctuation mode below. 

The wall thickness is roughly evaluated by the inverse of the 
curvature scale of the potential as $M^{-1}$. 
Since we are now assuming that $M^2\gg H^2$, 
it will not be too crude to adopt the thin-wall approximation.
Then one can evaluate ${\cal N}_W$ as
\begin{eqnarray}
{\cal N}_W=&&H\int d\tau a_E(\tau)({d\sigma_B\over d\tau})^2
\approx RS_1\,;
\nonumber\\
\quad &&S_1:=H\int d\tau ({d\sigma_B\over d\tau})^2,
\end{eqnarray}
where $S_1$ is the tension of the wall.
If one uses the reduced Euclidean action written in terms of
$R$, $S_1$, and the potential energy density inside
the bubble $V_R$ (true vacuum) and outside the bubble $V_L$
(false vacuum), $R$ is determined in terms of the other parameters 
by minimizing the action as\cite{Parke,Bellido}
\begin{eqnarray}
R={3S_1\over\sqrt{(\Delta V+6\pi GS_1^2)^2+24\pi GV_RS_1^2}}\,,
\label{Radius}
\end{eqnarray}
where $\Delta V=V_L-V_R$.
Then the final result for the amplitude
of the wall fluctuation mode is
\begin{eqnarray}
|{\cal R}_W|^2
&=&{8\pi GV_R\over9S_1^2}
\sqrt{(\Delta V+6\pi GS_1^2)^2+24\pi GV_RS_1^2}
\nonumber\\
&=&{H_R^2\over8\pi GS_1^2}
\sqrt{\left(H_L^2-H_R^2+(4\pi GS_1)^2\right)^2
+(8\pi GS_1 H_R)^2}\,,
\label{calRW}
\end{eqnarray}
where $H_L$ and $H_R$ are 
the Hubble parameters in the regions L and R,
respectively. Note that although we assumed $(H_L-H_R)/H_R\ll1$,
the difference $H_L-H_R$ may not be negligible 
in the determination of the amplitude of the wall fluctuation mode.

Now we consider the CMB power spectrum due to the
wall fluctuation mode. 
Fig.~4 shows the power spectrum of CMB anisotropies $l(l+1)C_l^{(W)}$
for various values of $\Omega_0$. The curves are normalized
by $H^2/RS_1$. As discussed before, the amplitude of the CMB
anisotropies due to the wall fluctuation mode are determined by the nature 
of the bubble wall.  This gives an independent constraint on open 
inflation models. In particular, the spectrum of CMB anisotropies
rises sharply towards lower multipoles for open models with
$\Omega_0\agt0.2$. Hence if this mode dominates over the other
mode contributions, such single field models will be severely excluded
 by the COBE data\cite{COBE}.
This issue has been recently investigated by Garriga \cite{Garriga} and
Garcia-Bellido\cite{Bellido}.

To carry on a further analysis, we have to specify the model in more
details.  Let us parametrize the potential
$V(\sigma)$ by $M^2$, $V_L$, $V_b$ and $\Delta V$ 
as shown in Fig.~1(a). We assume $V(\sigma)\ll M_{pl}^4$, $M\ll M_{pl}$
where $M_{pl}$ is the Planck mass, $\Delta V/V_L\ll1$ and
the potential has the unique curvature scale $M^2$ in the region C. 
Then, as the wall thickness is essentially given by $M^{-1}$,
we have
\begin{equation}
 S_1\sim {V_b\over M}\,.
\end{equation}
Further we assume $V(\sigma)\sim\lambda M_{pl}^4(\sigma/M_{pl})^{2n}$
($n=1,2,\cdots$) on the right of the barrier, where $\lambda\ll1$
as in the chaotic inflation.

There are three requirements to be satisfied. 
1) The tunneling must be dominated not by the 
Hawking-Moss instanton but by the Coleman-De Luccia instanton. 
This requires that $M^2>4H^2$. 
2) The tunneling rate $\Gamma$ must be suppressed enough in order to 
avoid the fluctuations caused by the bubble collisions. 
This implies that \cite{Col,DeLCol}
\begin{equation}
 -\ln(\Gamma/H^4)\sim {27\pi^2S_1^4\over2\Delta V^3}\gg 1.
\label{req2}
\end{equation}
3) Finally, the wall fluctuation mode must not dominate ${\cal R}_c$. 
Noting that the contribution of the continuous modes to 
its power in the logarithmic interval of $p$ at $p\gg1$
is $\langle{\cal R}_c^2\rangle:=
\lim_{p\to\infty}|{\cal R}_p|^2p^3/(2\pi^2)$, this requires
\begin{equation}
\vert{\cal R}_W\vert^2 \alt\langle{\cal R}_c^2\rangle
= \left({H_R^2\over2\pi\dot\sigma_B}\right)^2\,.
\label{req3}
\end{equation}
We assume the requirement 1) is satisfied 
and investigate the conditions on the potential parameters
derived from the requirements 2) and 3).

{}From the slow roll equation of motion,
we have $|H_R^2/\dot\sigma_B|
=8\pi H_RV_R/|V'(\sigma_B)|=8\pi\zeta H_R/M_{pl}$, where $\zeta\agt1$.
Therefore
\begin{equation}
{|{\cal R}_W|^2\over\langle{\cal R}_c^2\rangle}
={\pi\over8\zeta^2}\sqrt{\left(1+{\Delta V\over6\pi GS_1^2}\right)^2
+\left({H_R\over2\pi GS_1}\right)^2}\,.
\end{equation}
Since the left hand side of this must be smaller than unity,
we should have
\begin{equation}
{\Delta V\over6\pi GS_1^2}\alt\zeta^2\quad\hbox{and}\quad
{H_R\over2\pi GS_1}\alt\zeta^2\,.
\end{equation}
Using the fact $S_1\sim V_b/M$ and $\Delta V\ll V_L$, we find
the first inequality is automatically satisfied if the second one
is, which is re-expressed as
\begin{equation}
{V_b\over V_L}\agt{H_RMM_{pl}^2\over2\pi\zeta^2V_L}
={4\over3\zeta^2}\left({H_R\over H_L}\right){M\over H_L}\,.
\label{Vbcond}
\end{equation}
If this condition is satisfied, we also find the requirement 2) is 
fulfilled. Thus if the potential barrier is high enough,
approximately of $O(M/H_L)\times V_L$, the wall fluctuation mode
will become harmless. Apart from the intrinsic unnaturalness
of a single field model, it is not difficult to construct models
which satisfy the above constraint. Hence we conclude that
a single field model remains still viable.

\subsection{Two-field model}
Next we consider a two-field model 
in which the tunneling field $\sigma$ and 
the inflaton in the nucleated bubble $\phi$ are different.
The supernatural inflation model proposed by Linde and
 Mezhlumian\cite{Lindeb} is included in this category.
We consider the following situation. 
Before tunneling, $\phi$ 
is at the minimum of the potential $\phi_L$ with mass $M$. 
During the tunneling, the potential of $\phi$ 
changes to an almost flat but slightly declined one, i.e., 
$V_{\phi}'\ne 0$ and  $V_{\phi}''\sim 0$. 
Thus $\phi_L$ is no longer the minimum of the potential 
in the nucleated bubble.
Therefore $\phi$ begins the slow rolling to the new 
minimum of the potential. 

In the following calculation, 
we assume that the gradient of the potential  
is so small that we can 
neglect its effect in estimating the amplitude of 
the fluctuations $\varphi$, though
the condition under which this neglection is justified 
is not clear. 
Then our present formalism can be applied. 
For simplicity, 
we also assume that the thin wall approximation is valid.
We denote the radius of bubble wall by 
$R=a_E(\tau_0)=H^{-1}\cos \tau_0$.
Then the bubble wall trajectory in the Lorentzian region is described 
by the hypersurface $t_C=\tau_0={\rm const.}$ in the region C.

Under these assumptions, we can evaluate the fluctuation spectrum
inside the bubble by applying the formulas obtained in section 3
with ${\cal M}^2$ given by
\begin{equation}
 {\cal M}^2=\left\{
  \begin{array}{cc}
     0;&\quad (\tau_0<t_C<\pi/2), \\
     M^2;&\quad (-\pi/2<t_C<\tau_0).
  \end{array}
 \right.
\end{equation}
The form of the potential $U(\xi)$ in Eq.~(\ref{eqinCe})
in this case is illustrated in Fig.~5.
Thus the parameters of the present open inflationary model
are the mass of the inflaton field in the false vacuum $M$, the
wall radius $R$ and the Hubble parameter $H$.

First we consider the continuous spectrum.
As the mass is constant both outside and inside the wall, 
we find that the solution of the Eq.~(\ref{eqinCe}) becomes 
\begin{eqnarray}
\chi_{p}^{(R)}(t_C) & = & P^{ip}_{\nu'}(\sin t_C-i0); 
  \quad (\tau_0<t_C<\pi/2)
\nonumber \\
\chi_{p}^{(L)}(t_L) & = & P^{ip}_{\mu'}(\sin t_C-i0);
  \quad (-\pi/2<t_C<\tau_0).
\end{eqnarray}
The coefficients $\alpha_p$ and $\beta_p$ are determined by the junction
condition at the wall in the region C:
\begin{equation}
\left(\begin{array}{c}
\vspace{2mm} \chi^{(R)}_p \\ 
 \displaystyle {d \chi^{(R)}_p \over d\xi}\end{array}\right)
=\left(\begin{array}{cc}
\vspace{2mm} \chi^{(L)}_{-p} & \chi^{(L)}_{p} \\ 
 \displaystyle {d \chi^{(L)}_{-p}\over d\xi} 
 & \displaystyle
 {d \chi^{(L)}_{p} \over d\xi} \end{array}\right)
\left(\begin{array}{c}
 \alpha_p \\ ~ \\
 \beta_p \end{array}\right). 
\end{equation}
Then $\alpha_p$ and $\beta_p$ given by Eqs.~(\ref{alphbeta})
can be now easily evaluated on the wall. 

We have numerically evaluated the spectrum of the continuous modes.
Figs.~6 show the power spectra of $\bigl|{\cal R}_p\bigr|^2$
for various values of $M$ and $R$, normalized by the
 Bunch-Davies vacuum spectrum. 
Fig.~6(a) shows the case when the mass outside the bubble 
is $M/H=2$. The lines are the power spectra for
the wall radii $HR=0.1$, $0.5$, $0.7$, and $0.9$.
One sees the spectra for all the wall radii are almost the same
and they coincide with that of the Bunch-Davies vacuum
except for the small range of very small $p$. 
On the other hand, Fig.~6(b) is the case $M/H=10$
with the lines showing the power spectra for
$HR=0.1$, $0.5$, $0.7$, and $0.9$.
For $M/H\gg1$, we can apply the discussion given in
section 4A, around Eq.~(\ref{Mspec}). There we have seen that the
spectrum does not differ much from the case of the 
Bunch-Davies vacuum. Thus the increase in the amplitude
at $p\alt1$ will saturate as $M/H\to\infty$.
In both Figs.~6(a) and (b), all the lines rapidly 
approaches unity at $p\gg1$, which corresponds to the 
Harrison-Zel'dovich spectrum, in accordance with our expectation.

Next we turn to the discrete spectrum. 
Again the mode function is readily solved by means of the 
associated Legendre funciton. 
Thus eigenvalue probem reduces to Eq.~(\ref{regcond}) with
$\alpha_\Lambda$ and $\beta_\Lambda$ given in terms of the
junction condition at $t_C=\tau_0$ as
\begin{equation}
\alpha_\Lambda=\left.{((P_{\mu'}^{-\Lambda},P_{\nu'}^{-\Lambda}))
\over((P_{\mu'}^{-\Lambda},P_{\mu'}^{\Lambda}))}
\right|_{t_C=\tau_0}\,,
\quad
\beta_\Lambda=\left.{((P_{\nu'}^{-\Lambda},P_{\mu'}^{\Lambda}))
\over((P_{\mu'}^{-\Lambda},P_{\mu'}^{\Lambda}))}
\right|_{t_C=\tau_0}\,.
\label{supjunct}
\end{equation}

It is instructive to show an analytically
solvable example of this eigenvalue problem.
Let us consider the case $M^2=2H^2$ ($\mu'=0$) and $m^2=0$
($\nu'=1$). 
In this case, the associated Legendre functions are
expressed in terms of elementary functions. 
\begin{eqnarray}
  &&P^{\pm\Lambda}_0(z)={1\over\Gamma(1\mp\Lambda)}
  \biggl({z+1\over z-1}\biggr)^{\pm\Lambda/2},
\nonumber\\
  &&P^{\pm\Lambda}_1(z)={z\mp\Lambda\over\Gamma(2\mp\Lambda)}
  \biggl({z+1\over z-1}\biggr)^{\pm\Lambda/2},
\end{eqnarray}
Also, Eq.~(\ref{regcond})
reduces simply to $\beta_\Lambda=0$,
hence from Eq.~(\ref{supjunct}),
\begin{equation}
  ((P_1^{-\Lambda},P_0^\Lambda))=0.
\end{equation}
This is easily solved to give
\begin{equation}
  \Lambda={-x_0+\sqrt{2-x_0^2}\over2},
\end{equation}
where $x_0=\sin\tau_0=\sqrt{1-(HR)^2}$.
This result shows that $\Lambda$ becomes larger
as the wall radius increases.

{}For general $M^2$, we need numerical evaluation. 
Since $0<\Lambda\leq1$ in the exact 
de Sitter case with the upper limit attained when $m^2\to0$, 
we expect any introduction of finite mass outside the wall will reduce 
$\Lambda$ to a value
less than unity. Hence we look for an eigenvalue in the range
$0<\Lambda<1$.
Fig.~7 shows $\Lambda$ as a function of $HR$
for several values of $M^2$ when $m^2=0$.
One sees $\Lambda$ becomes smaller as $M^2$ becomes larger 
or as the wall radius becomes smaller.
But the de Sitter super-curvature mode exists for
any wall radius as long as $M^2<2H^2$. 
On the other hand, for $M^2>2H^2$,
the de Sitter super-curvature mode
ceases to exist when
the wall radius becomes smaller than a critical value.
Fig.~8 shows the critical line on the $(M/H,HR)$-plane
on which the super-curvature mode disappears. 
The de Sitter super-curvature mode exists below the line.

This property of the super-curvature mode can be understood
in analogy with the quantum mechanics described before. 
As the wall radius decreases, the position of
the bubble wall $\xi_0$ moves right in Fig.~5 of the potential 
$U(\xi)$. Also, as the mass $M$ increases, the
potential barrier outside the bubble wall $\xi<\xi_0$
becomes higher. These make it difficult 
to form a bound state.
We note that $U(\xi)$ has the deepest valley when
 the mass $M$ is zero.
This situation corresponds to the case when we assume
the Bunch-Davies vacuum state inside the bubble and
when the super-curvature mode contributes extremely
\cite{YST95}.

Now we show the results of numerical calculations of 
the CMB anisotropy power spectrum $l(l+1)C_l$ in Fig.~9
for the $\Omega_0=0.1$ universe
 with $M^2/H^2=2$ and the wall radii $HR=0.3$, $0.5$, $0.7$, and $0.9$.
The curves are normalized by $(3H^2/5\dot\phi_B)^2$.
Here both continuous and discrete modes are included. 
To compare the results with the previous ones
\cite{LS,RPb,Kamion,Gorski,YST95,YB},
we plotted two dashed lines in the figure.
The top dashed curve is the result when the scalar field is in the 
Bunch-Davies vacuum \cite{YST95,YB}, and the bottom dashed curve
is the one in the conformal vacuum\cite{LS,RPb,Kamion,Gorski}.
All of the curves lie between these two curves.
We see the amplitude becomes large as the wall radius increases,
approaching that in the case of the Bunch-Davies vacuum.
As described before, the contribution of the
super-curvature mode is most for the Bunch-Davies vacuum
 and is least for the conformal vacuum.
When the wall radius becomes large and
the mass becomes small, the contribution from the 
super-curvature mode becomes large, which explains the behavior
of the CMB power spectra in Fig~.9.
We have also calculated the CMB spectra for the $\Omega_0=0.3$
universe. The results have turned out
to be almost independent of the model parameters.
In fact, if only $C_l^{(C)}$ is taken into account,
the difference between 
the Bunch-Davies vacuum and the conformal vacuum is only a
few percent even for $\Omega_0=0.1$ \cite{YST95}.
Thus the differences are dominantly due to the de Sitter super-curvature
mode $C_l^{(S)}$, but its contribution rapidly becomes negligible as 
one increases $\Omega_0$.

\section{Conclusion}

 We have investigated in detail the quantum fluctuations and
the resultant CMB anisotropies in two simple
models of one-bubble open inflationary scenario.
Before going into the analysis of the specific models, we have
extended the previously developed formalism to investigate the 
quantum state inside the nucleated bubble, and derived 
formulas for the mode functions which take account of 
the effect of the tunneling described by the 
Coleman-De Luccia instanton. 
In the main stream, 
we have assumed the potential energy difference 
between the false and true vacua is small enough so that 
spacetime both inside and outside the bubble
can be described by de Sitter space with a single Hubble parameter $H$, 
but the results can be easily extended to more general cases. 

A complete description of a quantum state requires a complete set of 
mode functions which are normalizable on a Cauchy surface of the whole 
spacetime. 
This brings about new sets of fluctuation modes which had not been
considered in most of the previous analyses of an open inflationary
universe. In addition to the usual modes with continuous 
spectrum, there are modes with discrete spectrum corresponding 
to the fluctuations on super-curvature scales. 
For both the continuous and discrete modes, 
we have obtained general formulas of the spectrum of 
the CMB anisotropies on large angular scales which result from
the quantum fluctuations.

After these preparations, we have considered two simple 
classes of one-bubble open inflationary models. 
One is a single field model with the potential
as illustrated in Fig.~1(a), assuming that the mass square
in the false vacuum, $M^2$, is much larger than $H^2$. 
The other is a two-field model in which the false vacuum decay is
mediated by a scalar field different from the inflaton inside the bubble
and the inflaton is massive in the false vacuum
through the coupling with the tunneling field.

In the case of a single field model with $M^2\gg H^2$,
there is one discrete mode which represents fluctuations of the bubble
wall. We have shown that there is 
no other discrete mode in this model. 
Thus we have considered the 
CMB anisotropies due to the continous modes and the 
wall fluctuation mode. 
As far as the contribution of the continuous modes to the CMB 
anisotropies is concerned, we have found that it is approximately
the same as in the case of the Bunch-Davies vacuum or
the conformal vacuum. Hence we have focused on the wall
fluctuation mode.
The curvature perturbation due to the wall fluctuation mode is notable.
It is always present in this model 
and its amplitude is totally determined by the part of the 
potential which governs the tunneling but has nothing to do 
with the scalar field dynamics inside the nucleated bubble.
Since its amplitude is independent of the amplitude of the curvature
perturbation due to the continuous modes,
its contribution to the CMB anisotropies gives rise to an independent
constraint on the model, just as the gravitational wave perturbation
does to the usual inflation models. Fortunately, the constraint turns 
out to be relatively weak. 
Thus it is possible to construct a model in which 
the effect of the wall fluctuation mode can be neglected. 

Here we make the following remark. 
The wall fluctuation mode 
has a peculiar property that curvature perturbations 
induced by it are transverse-traceless, 
hence it can be regarded as a mode of tensor-type perturbations. 
This means that the perturbation of the scalar field 
might couple with the gravitational wave perturbation for this mode 
when the degrees of freedom of gravity are fully taken into accout. 
But we have not incoorpolated them in the present analysis. 
Thus there remains a possibility that 
the final answer changes qualitatively.
We hope to come back to this issue in near future. 

As for a two-field model, we have adopted the thin-wall 
approximation and assumed the mass of the inflaton 
changes like the step function across the wall.
Thus the model can be parametrized by the mass in the 
false vacuum $M$, the Hubble radius $H$ and the wall radius $R$.
In this model, 
the origin of the discrete mode is different from that in 
the single field model. 
Since it originates from the spacetime structure of de Sitter space, 
we have called it the de Sitter super-curvature mode. 
The existence of the de Sitter super-curvature mode depends
on the mass $M$ of the inflaton at the false vacuum and
on the radius $R$ of the bubble wall.
It appears when $HR$ is large and $M/H$ is small.
For models with $M/H \gg 1$, the super-curvature mode disappears.
Different from the wall fluctuation mode which appeared in the 
single field model, 
the amplitude of spatial curvature perturbations induced by this mode
is determined by the scalar field dynamics
at the open inflationary stage inside the bubble,
as in the case of the continuous modes. 

We have then investigated the spectrum of CMB anisotropies on
large angular scales in this model.
Though the spectra of the curvature perturbations due to
the continuous modes have different shapes 
on the curvature scale for different model parameters, we have found 
the resulting CMB power spectra do not significantly depend on 
the parameters for $\Omega_0\agt0.1$.  
On the other hand, the parameter-dependence appears clearly
in the contribution of the de Sitter super-curvature
mode to the CMB spectrum.
The effect of this mode is appreciable when
the wall radius is large ($HR\sim1$) and the mass is small ($M/H\ll1$).
The effect is to raise the amplitudes of low multipoles at $l\alt10$
for models with $\Omega_0\alt0.1$.
If this mode contributes significantly, the predicted CMB power
spectra will contradict with COBE observations \cite{YB}.
However, for an open universe of $\Omega_0\agt0.3$ as well as
for models with small $HR$ and large $M/H$,
the effect of the super-curvature mode on the CMB anisotropies
is practically negligible. 

We comment on implications of our results to the large scale
structure formation. As all our open models predict
the Harrison-Zel'dovich spectrum on small scales,
the difference can appear only from
the normalization of the density perturbations.
However, the difference will be negligible if we
adopt the normalization scheme in terms of the
likelihood analysis using the COBE result\cite{COBE}.
This is because all the models predict practically the same CMB power 
spectrum at $l\agt10$.

In summary,
 we have presented a detailed analysis of the quantum fluctuations
in open universe in simple models of the one-bubble inflationary
 scenario and the resulting CMB anisotropies on large angular scales.
We have found there exist ranges of model 
 parameters which are consistent with CMB observations.
Thus the one-bubble inflationary scenario is a viable one
for explaining the large scale structure of the universe
and it certainly deserves further study.
There are of course many issues left to be clarified in future.
For example, consideration of the continuous gravitational wave
modes is definitely necessary. Analyses of more sophisticated models,
like several other two-field models proposed by Linde and
Mezhlumian\cite{Lindea,Lindeb} are of particular interest.
Inclusion of the degrees of freedom of the gravitational perturbation in 
the formalism from the beginning is a difficult issue but should
be done in order to gain a more firm picture of the false vacuum 
decay and the subsequent evolution of the quantum state.

\vspace{5mm}
\begin{center}
{\bf Acknowledgements}
\end{center}
We thank N. Sugiyama and H. Ishihara for discussions.
We also thank J. Garcia-Bellido for
showing us his result prior to publication.
\vspace{1cm}

\begin{appendix}
\section{Derivation of the evolution equation for ${\cal R}_c$}

Here we derive the evolution equation for ${\cal R}_c$. We follow the
notation of Ref.~\cite{KS} for the perturbation variables.
Consider the general scalar harmonics on 3-space with curvature constant $K$,
which satisfy
\begin{equation}
({\mathop\Delta^{(3)}} +k^2)Y=0.
\end{equation}
The corresponding vector and tensor harmonics are defined as
\begin{equation}
Y_{i}=-{1\over k}Y_{|i}\,,\quad 
Y_{ij}={1\over k^{2}}Y_{|ij}+{1\over3}\gamma_{ij}Y,
\end{equation}
where the vertical bar denotes the covariant derivative with respect
to the 3-metric $\gamma_{ij}$.
In the notation of the present paper, we have set $K=-1$ and $k^2=p^2+1$.
But since we do not have to fix $K$ in the discussion below,
we leave it arbitrary and use $k$ to denote the eigenvalue.
Also we suppress the eigenvalue index $k$ for
notational simplicity.
We expand all the perturbation variables in terms of these harmonics.
Specifically, the metric is expressed as
\begin{equation}
ds^2=a^2\left[-(1+2AY)d\eta^2-2BY_idx^id\eta
+\left((1+2H_L)\gamma_{ij}+2H_TY_{ij}\right)dx^idx^j\right],
\end{equation}
and the energy-momentum tensor is expressed as
\begin{eqnarray}
&&T^0{}_0=-\rho(1+\delta\, Y)\,,\quad T^i{}_0=-(\rho+p)vY^i,
\nonumber\\
&&T^i{}_j=p\left((1+\pi_LY)\delta^i_j+\pi_TY^i{}_j\right).
\end{eqnarray}
The background equations are
\begin{equation}
\left({a'\over a}\right)^2+K={8\pi G\over3}\rho a^2\,,
\quad \rho'+3{a'\over a}(\rho+p)=0\,,
\end{equation}
where the prime denotes the derivative with respect to $\eta$.
Then the Einstein equations for scalar-type perturbations are written as
\begin{eqnarray}
&&\delta G^0{}_0=-8\pi G\rho\delta\, Y\,,
\nonumber\\
&&\delta G^0{}_i=8\pi G(\rho+p)(v-B)Y_i\,,
\nonumber\\
&&\delta G^i{}_j=8\pi G(p\pi_LY\gamma_{ij}+p\pi_TY_{ij})\,,
\end{eqnarray}
where the explicit expressions for 
$\delta G^\mu{}_\nu$ can be found in Appendix D of Ref.~\cite{KS}. 

For our discussion, we need the $(0,i)$-component and the traceless part
of $(i,j)$-component of the above equations. They give, respectively,
\begin{eqnarray}
&&k{a'\over a}A-k{\cal R}'+K\sigma_g={4\pi G}(\rho+p)a^2(v-B)\,,
\label{0icomp}\\
&&k^2(A+{\cal R})-{1\over a^2}(a^2k\sigma_g)'=-8\pi Gpa^2\pi_T\,,
\label{ijcomp}
\end{eqnarray}
where ${\cal R}=H_L+{1\over3}H_T$ and $\sigma_g=k^{-1}H'_T-B$.
Expressing Eq.~(\ref{0icomp}) on the Newtonian hypersurface (defined by
$\sigma_g=0$), for which we have $A=\Psi$, ${\cal R}=\Phi$, $v-B=V$,
we find
\begin{equation}
\Phi'={a'\over a}\Psi-{4\pi G}(\rho+p)a^2{V\over k}\,.
\label{Phieq}
\end{equation}
Also, from Eq.~(\ref{ijcomp}), we obtain
\begin{equation}
\Psi+\Phi=-{8\pi G}{a^2\over k^2}p\Pi\,,
\label{PhiPsi}
\end{equation}
where $\Pi=\pi_T$ is the gauge-invariant anisotropic stress
 perturbation.

Using the contracted Bianchi identities,
which give the energy-momentum conservation law $T_\mu{}^{\nu}{}_{;\nu}=0$,
one obtains the equations for the matter variables.
Here we only need the $\mu=i$ component of it, i.e., the
equation for $V$. It is given by Eq.~$(4.7b)'$
in Chapter II of \cite{KS}, which is 
\begin{equation}
{1\over a}\left({aV\over k}\right)'=\Psi+S\,;
\quad S:={c_s^2\Delta+\Gamma\over1+w}
-{2\over3}{k^2-3K\over k^2}{w\over1+w}\Pi\,,
\label{Veq}
\end{equation}
where $c_s^2=p'/\rho'$, $w=p/\rho$,
$\Gamma$ is the gauge-invariant entropy perturbation, and
$\Delta$ is the density perturbation on the comoving hypersurface
which is related to that on the Newtonian hypersurface $\Delta_s$ as
\begin{equation}
\Delta=\Delta_s+3(1+w){a'\over a}{V\over k}\,.
\end{equation}

Now from the gauge transformation property of ${\cal R}$,
the curvature perturbation ${\cal R}_c$
on the comoving hypersurface (defined by $v-B=0$) is expressed 
in terms of $\Phi$ and $V$ as
\begin{equation}
{\cal R}_c=\Phi-{a'\over a}{V\over k}\,.
\label{Rcdef}
\end{equation}
Then from Eqs.~(\ref{Phieq}) and (\ref{Veq}), we find
\begin{equation}
{\cal R}_c'=-K{V\over k}-{a'\over a}S\,.
\label{Rcprime}
\end{equation}

Taking the derivative of this equation and using Eqs.~(\ref{PhiPsi}) and
(\ref{Veq}), we finally obtain
\begin{equation}
{\cal R}_c''+2{a'\over a}{\cal R}_c'-K{\cal R}_c
=-\left(2\left({a'\over a}\right)^2+K\right)S-\left({a'\over a}S\right)'
+{K\over k^2}8\pi Gpa^2\Pi\,.
\end{equation}
We find the right hand side of this equation vanishes
if the universe is matter-dominated ($p=c_s^2=0$).
Furthermore, it is known that $\Delta=O\bigl((ka/a')^2\bigr)$ on
superhorizon scales\cite{KS}. From this fact, one can easily deduce 
${\cal R}_c$ remains constant on super-horizon scales if $\Gamma=\Pi=0$.
We also note that from Eqs.~(\ref{Rcdef}) and (\ref{Rcprime}),
$\Phi$ can be expressed as
\begin{equation}
\Phi={\cal R}_c-{1\over K}{a'\over a}({\cal R}_c'+{a'\over a}S)\,.
\end{equation}
In particular, if $S=0$ as in the case of a matter-dominated universe,
we have
\begin{equation}
\Phi={\cal R}_c-{1\over K}{a'\over a}{\cal R}_c'\,.
\label{PhiRc}
\end{equation}
With $K=-1$, this is the relation we have used in the text.

\section{Equavalence of tensor and scalar treatments for {$p^2=-4$}}
Here we show that the wall fluctuation mode can be treated 
either as a tensor-type perturbation or a scalar-type perturbation
and prove the equivalence of the CMB anisotropy formulae
in both approaches.

In general, setting $\delta T/T=\Theta$, the perturbed collisionless
Boltzmann equation is written as
\begin{equation}
{D\over d\lambda}\Theta
=kAY_i\gamma^i-({\cal R}'-{1\over3}k\sigma_g)Y
-k\sigma_gY_{ij}\gamma^i\gamma^j\,,
\label{deltaT}
\end{equation}
where $D/d\lambda$ denotes the Lagrange derivative along the light ray
with $\lambda$ being the conformal affine parameter, and
$\gamma_i$ is the directional cosine of the photon propagation
vector.
Assuming the universe is matter-dominated,
one has $A=0$ on the comoving hypersurface $v-B=0$. 
Also, expressing Eq.~(\ref{0icomp}) on the comoving hypersurface, 
one obtains
\begin{equation}
{\cal R}_c'={K\over k^2}(k\sigma_g)_c\,,
\end{equation}
where $(k\sigma_g)_c$ is $k\sigma_g$ evaluated on the 
comoving hypersurface.
Using these facts, Eq.~(\ref{deltaT}) on the comoving hypersurface
becomes
\begin{equation}
{D\over d\lambda}\Theta_c=
{k^2-3K\over3k^2}(k\sigma_g)_cY-(k\sigma_g)_cY_{ij}\gamma^i\gamma^j\,.
\end{equation}
In particular, for $k^2=3K$ which corresponds to $p^2=k^2-1=-4$
for $K=-1$, one has
\begin{equation}
{D\over d\lambda}\Theta_c
=-(k\sigma_g)_cY_{ij}\gamma^i\gamma^j
=-3{\cal R}_c'Y_{ij}\gamma^i\gamma^j\,.
\end{equation}
Noting that $Y_{ij}$ in this case is transverse-traceless\cite{HAMA},
we see this is equivalent to the tensor-type perturbation
with the metric perturbation $H_{ij}=6{\cal R}_cY_{ij}$.

Now we go to the Newtonian slicing, in which $\sigma_g=0$,
by the transformation $\eta\to\eta+T$.
Then the gauge function $T$ is determined from
\begin{equation}
0=(k\sigma_g)_c-k^2T\,,
\end{equation}
which gives
\begin{equation}
T={(k\sigma_g)_c\over k^2}={{\cal R}_c'\over K}\,.
\label{gauge}
\end{equation}
Let $\Theta_s$ denotes the temperature anisotropy
on the Newtonian hypersurface.
The gauge transformation affects only the monopole and dipole parts
of the anisotropy. Then one finds $\Theta_s$ is related to
$\Theta_c$ as
\begin{equation}
\Theta_s=\Theta_c+{a'\over a}TY+TY_{|i}\gamma^i\,.
\end{equation}
Inserting Eq.~(\ref{gauge}) into this equation, we obtain
\begin{equation}
\bigl[\Theta_c\bigr]^{\eta_{0}}_{\eta_{LSS}}
=\bigl[\Theta_s-{a'\over a}{1\over K}{\cal R}_c'Y
-{1\over K}{\cal R}_c'Y_{|i}\gamma^i\bigr]^{\eta_{0}}_{\eta_{LSS}}\,,
\label{Trel}
\end{equation}
where $[\cdots]^{\eta_0}_{\eta_{LSS}}$ denotes the difference
between the quantity evaluated at $\eta_0$ and $\eta_{LSS}$.

On the other hand, the equation for $\Theta_s$ is obtained from
Eq.~(\ref{deltaT}) by setting $\sigma_g=0$ as
\begin{equation}
{D\over d\lambda}\Theta_s=\Phi Y_{|i}\gamma^i-\Phi'Y
={D\over d\lambda}(\Phi Y)-2\Phi'Y\,,
\label{Tseq}
\end{equation}
where we have used the fact $\Phi+\Psi=0$, which follows from
Eq.~(\ref{PhiPsi}).
Combining Eqs.~(\ref{Trel}) and (\ref{Tseq}), 
we find
\begin{equation}
\bigl[\Theta_c\bigr]^{\eta_{0}}_{\eta_{LSS}}
=\bigl[\Phi-{a'\over a}{1\over K}{\cal R}_c'Y
-{1\over K}{\cal R}_c'Y_{|i}\gamma^i\bigr]^{\eta_{0}}_{\eta_{LSS}}
-2\int^{\eta_0}_{\eta_{LSS}}\Phi'Yd\lambda\,.
\end{equation}
Noting that we have 
\begin{equation}
\Phi\to{3\over5}{\cal R}_c\,,\quad 
{a'\over a}{\cal R}_c'\to-{2\over5}{\cal R}_c\,,
\quad {\cal R}_c'\to0\,,
\end{equation}
in the limit $\eta\to0$ (with $K=-1$),
the above equation reduces to
\begin{equation}
\bigl[\Theta_c\bigr]^{\eta_{0}}_{\eta_{LSS}}
=-{1\over3}\Phi Y\Bigr\vert_{LSS}
-2\int^{\eta_0}_{\eta_{LSS}}\Phi'Yd\lambda
+(\hbox{monopole}+\hbox{dipole})\,,
\end{equation}
for $\eta_{LSS}\to0$. This is just the formula to
evaluate the CMB anisotropies due to scalar-type perturbations\cite{GSS}.
Thus we have proved the equivalence of the tensor and scalar approaches
to the CMB anisotropies when $p^2=-4$, i.e., for the wall fluctuation mode.

\section{Absence of the de Sitter 
super-curvature mode in the single field model}

Here we show there exists no discrete mode 
other than the wall fluctuation mode in the 
single field model, provided the wall thickness, i.e.,
the inverse mass scale of the potential is much smaller 
than the Hubble radius $H^{-1}$ as well as than the wall radius $R$. 
We consider the following simplified model in which 
the potential in Eq.~(\ref{eqinCe}) takes the form,
\begin{equation}
U(\xi)=\left\{\begin{array}{llc}
 U_{\rm I}\displaystyle{\cosh^2\xi_0\over \cosh^2 \xi}~& 
  \left(\xi<\xi_0 -\varepsilon\right);~&{\rm (I)}\,,\\
\vspace{2mm} -U_{\rm II} &
  \left(\xi_0-\varepsilon<\xi<\xi_0\right);~&{\rm (II)}\,,\\
  U_{\rm III} &
  \left(\xi_0<\xi<\xi_0+\varepsilon\right);~&{\rm (III)}\,,\\
 -\displaystyle{2\over \cosh^2 \xi} & 
  \left(\xi_0+\varepsilon <\xi<\xi_0+\varepsilon\right);~&{\rm (IV)}\,,
   \end{array}\right.
\end{equation}
where $U_{\rm I}:=(M^2/H^2-2)/\cosh^2 \xi_0$ (see Fig.~10).
We assume $U_{\rm I}$, $U_{\rm II}$, $U_{\rm III}\gg1$,
in accordance with the assumption $M^2/H^2\gg1$.
We note that $\xi_0$ must be positive, since $\dot a$ is negative inside
the wall.
We consider the solution $\chi_p$ with $p=i\Lambda$, where
$0<\Lambda\leq2$. Then we may approximate the potential in (I)
by the constant $U_{\rm I}$. 
Further, for simplicity, we set $U_{\rm I}=U_{\rm III}$.
Then the solution takes the form,
\begin{equation}
 \chi_\Lambda=\left\{\begin{array}{lc}
C_{\rm I} e^{\kappa (\xi-\xi_0)} & (\hbox{I}),\\
C_{\rm II} \left(e^{ik (\xi-\xi_0)}+b_{\rm II} 
       e^{-ik (\xi-\xi_0)}\right) & (\hbox{II}),\\
C_{\rm III} \left(e^{\kappa (\xi-\xi_0)} 
      +b_{\rm III} e^{-\kappa (\xi-\xi_0)}\right) & (\hbox{III}),\\
C_{\rm IV} e^{-\Lambda (\xi-\xi_0)}(\tanh\xi+\Lambda)~& (\hbox{IV}),
     \end{array}\right.
\end{equation}
where 
\begin{equation}
 \kappa =\sqrt{U_{\rm I}+\Lambda^2},\quad
 k=\sqrt{U_{\rm II}-\Lambda^2},
\end{equation}
and $C$'s and $b$'s are constants determined by the continuity 
conditions of $\chi_\Lambda$ and $d\chi_\Lambda/d\xi$.
In (I), we take only the growing solution ($\propto e^{\kappa\xi}$)
 because the contribution from the decaying solution
 ($\propto e^{-\kappa\xi}$) will be exponentially suppressed.

Solving the continuity condition of $d(\ln\chi_\Lambda)/d\xi$
 at $\xi=\xi_0-\varepsilon$ and $\xi=\xi_0$, 
we find $d(\ln\chi_\Lambda)/d\xi$ at ${\xi=\xi_0+\varepsilon}$
is calculated from the left of the point to be
\begin{equation}
  \psi_L:=\kappa\,{\sin(2\theta-k\varepsilon) e^{\kappa\varepsilon}-
         \sin \kappa\varepsilon\, e^{-\kappa\varepsilon} \over 
         \sin(2\theta-k\varepsilon) e^{\kappa\varepsilon}+
         \sin \kappa\varepsilon\, e^{-\kappa\varepsilon}}\,,
\label{juncL}
\end{equation}
where 
\begin{equation}
\tan\theta=\kappa/k\,.
\end{equation}
On the other hand, the same quantity is calculated from the
right of $\xi_0+\varepsilon$ to be
\begin{equation}
\psi_R: =-\Lambda +{1\over \cosh^2({\xi_0+\varepsilon})
     (\tanh({\xi_0+\varepsilon})+\Lambda)}\,.
\label{juncR}
\end{equation}
Thus the continuity condition at $\xi=\xi_0+\varepsilon$
reduces to
\begin{equation}
\psi_L=\psi_R\,.
\label{junct}
\end{equation}
This equation must be satisfied for $\Lambda=2$
because we know the wall fluctuation mode always exists. 
This condition gives one constraint on the model parameters 
$U_{\rm I}, U_{\rm II}$ and $\varepsilon$. 
Therefore all of them cannot be chosen arbitrarily.
This reflects the fact that ${\cal M}^2=V''(\sigma_B)$.
Suppose we fix $U_{\rm I}$ and $U_{\rm II}$, or equivalently, 
$\tilde\kappa:=\kappa\vert_{\Lambda=2}$ and 
$\tilde k:=k\vert_{\Lambda=2}$. 
Then $\varepsilon$ is determined from Eq.~(\ref{junct}) 
with $\Lambda=2$.
In order to find the order of magnitude of $\varepsilon$,
we note that $-2<\psi_R<-3/2$ when $\Lambda=2$,
i.e, $\psi_R=O(1)$.
Since $\kappa\gg 1$, this implies
the numerator in $\psi_L$ must be small,
\begin{equation}
\sin(2\tilde\theta-\tilde k\varepsilon)\, e^{\tilde\kappa\varepsilon}-
         \sin \tilde\kappa\varepsilon\, e^{-\tilde\kappa\varepsilon}\ll 1, 
\end{equation}
where $\tilde \theta=\theta\vert_{\Lambda=2}$.
Thus we find that $\varepsilon=O(1/\kappa)$. 
Using the above inequality, 
the denominator in $\psi_L$ is evaluated for $\Lambda=2$ as
\begin{equation}
 \sim 2\sin \tilde \kappa\varepsilon\, e^{-\tilde\kappa\varepsilon},
\end{equation}
and is of order unity. 

Now we show that no other discrete mode exists. 
Suppose there existed another discrete mode. 
The next largest eigenvalue solution must have one node. 
Noting the constraints $\xi_0>0$ and $\Lambda>0$, 
we can easily see that there would be no node in (IV). 
Thus the solution must have a node in (II) or (III). 
As we decrease $\Lambda$ from 2,
because $\chi_{\Lambda}$ must vanish at $\xi=\xi_0+\varepsilon$
just before a node would first appear in (III),
 $\psi_L$ would diverge for some value of $\Lambda$.
However, the changes of $k$ and $\kappa$ 
due to the variation of $\Lambda$ would be at most of 
$O(1/\tilde k)$ and $O(1/\tilde \kappa)$, respectively, 
and accordingly the change of $\theta$ would be also small.
Hence there is no chance that the denominator in $\psi_L$ would
vanish, which is a contradiction. 
Thus we conclude that there is no discrete mode 
other than the wall fluctuation mode in this 
limiting case.

\end{appendix}

\vspace{5mm}
\begin{center}
\underline{Figure Captions} 
\end{center}

\vspace{3mm}
\noindent
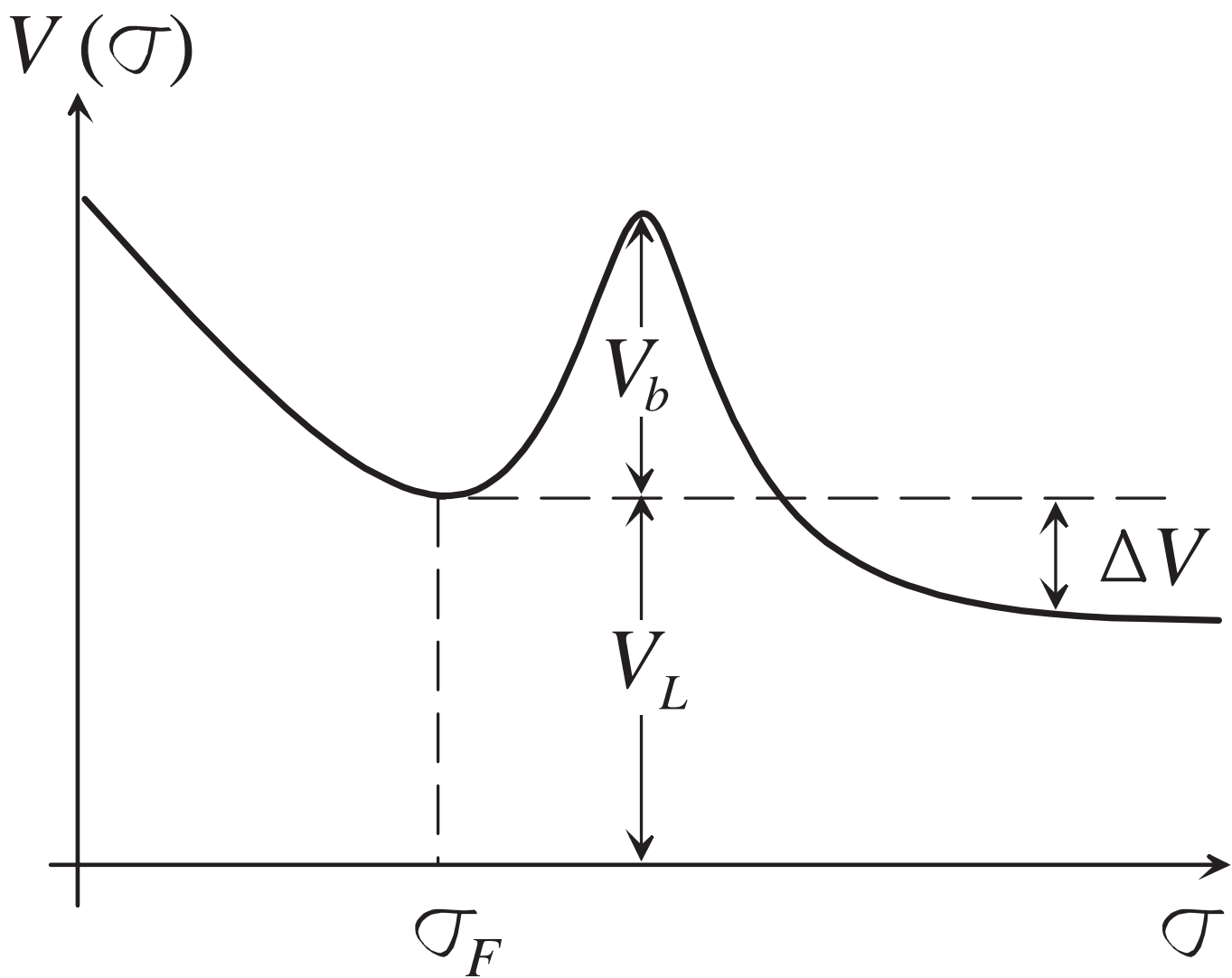
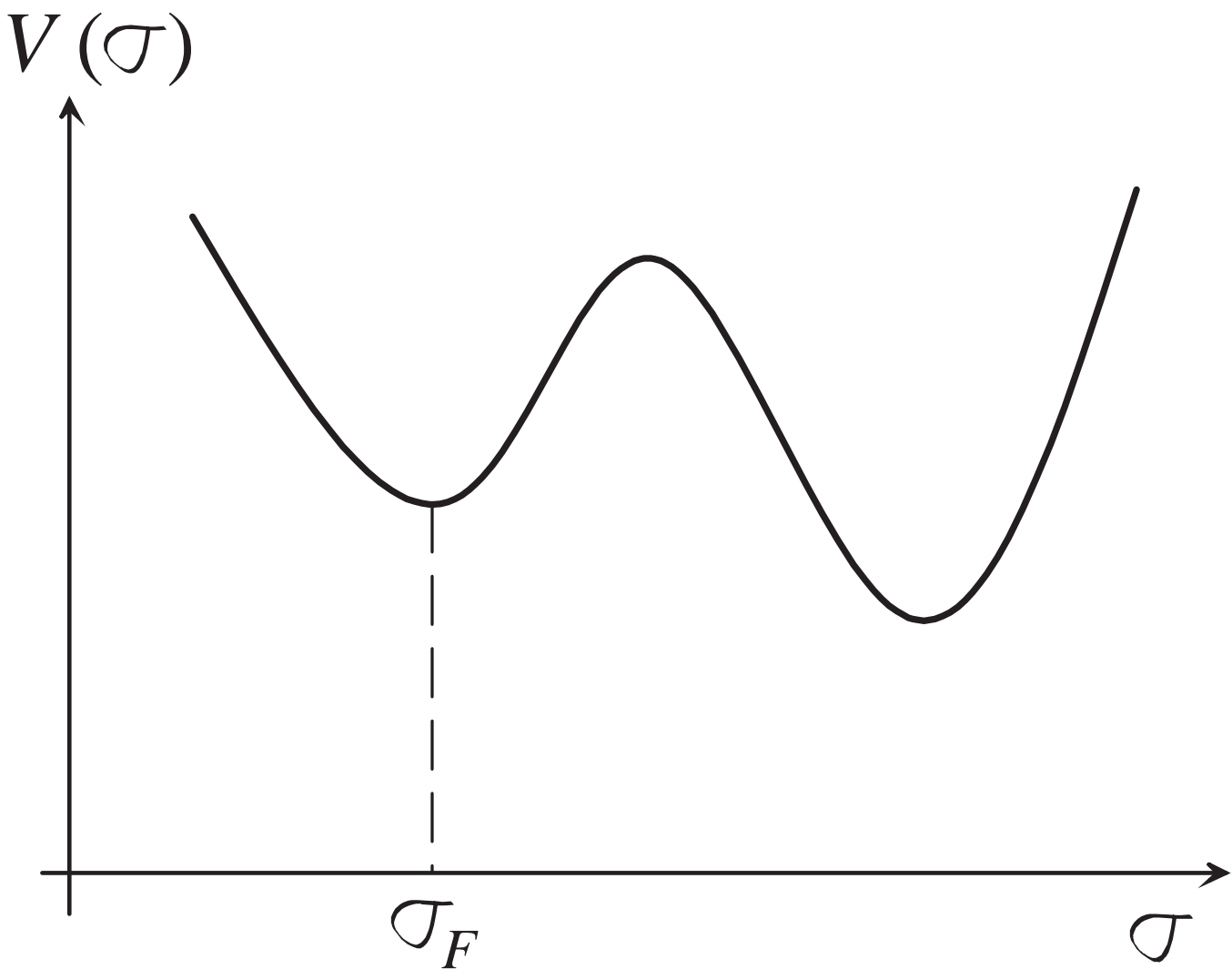
{Figure 1.}
A schematic picture of the tunneling potential 
for the one-bubble open inflationary scenario,
(a) for a single field model and (b) for
a two-field model.

\vspace{3mm}
\noindent
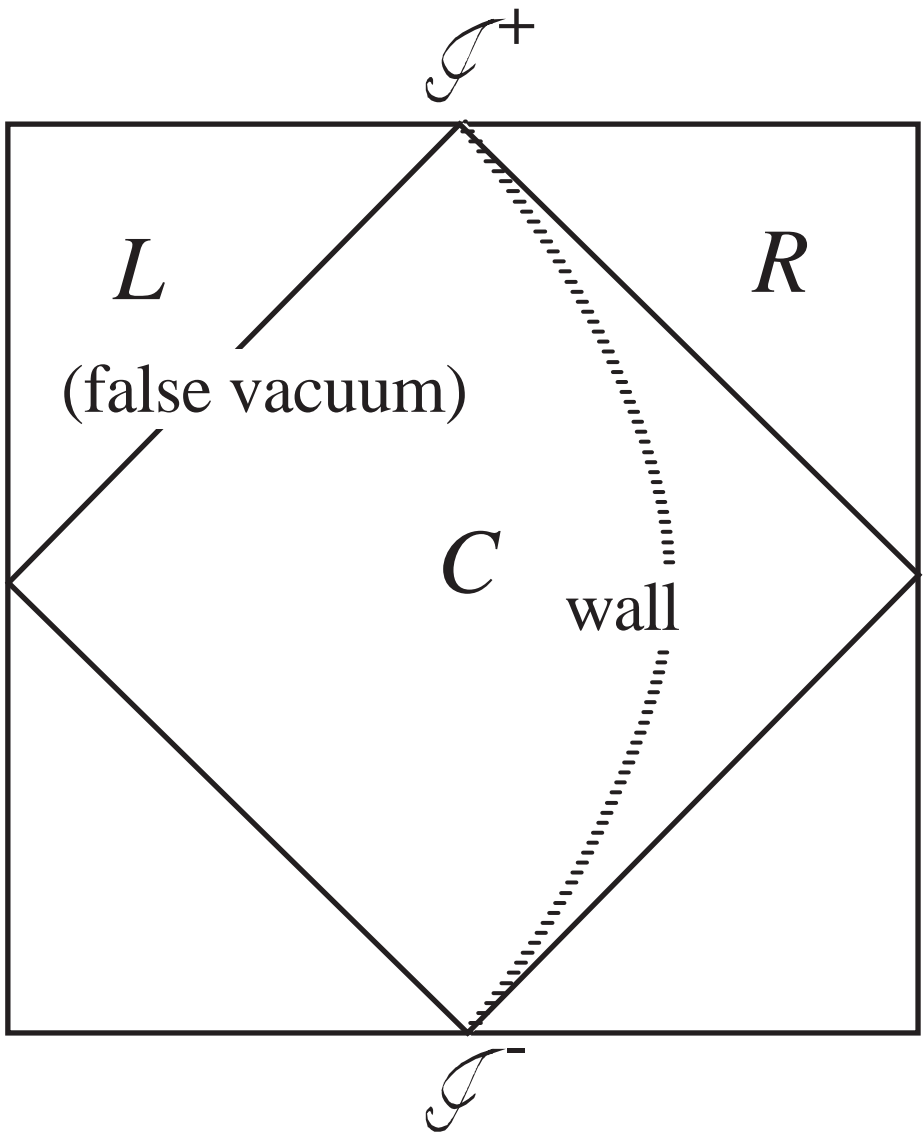
{Figure 2.} 
Penrose diagram of the de Sitter space.
The coordinates which cover the regions R, L and C
are shown, respectively.

\vspace{3mm}
\noindent
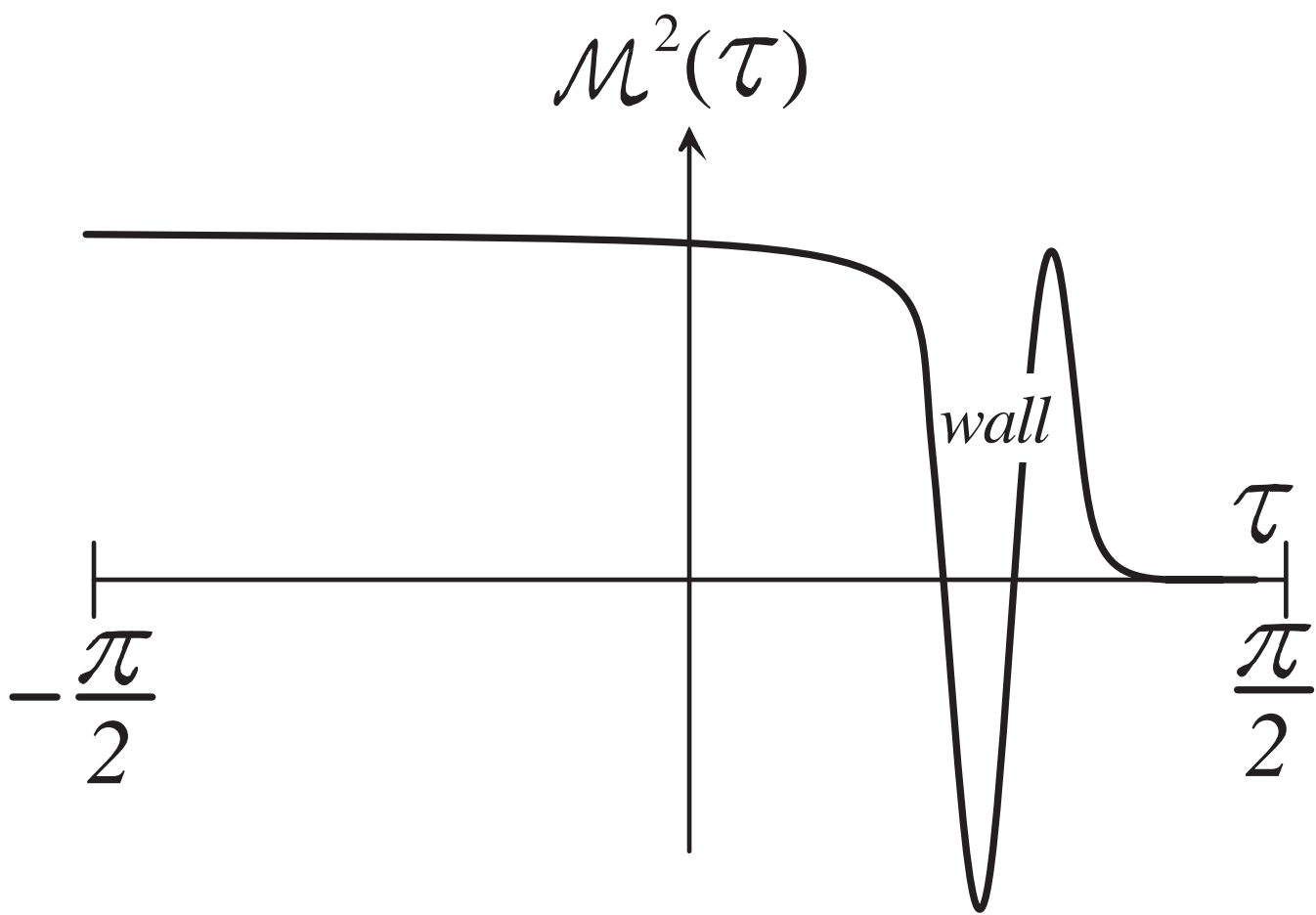
{Figure 3.}
A schematic picture of ${\cal M}^2$ as a function of
$\tau$ ($=t_C$) in the case of a single field model.

\vspace{3mm}
\noindent
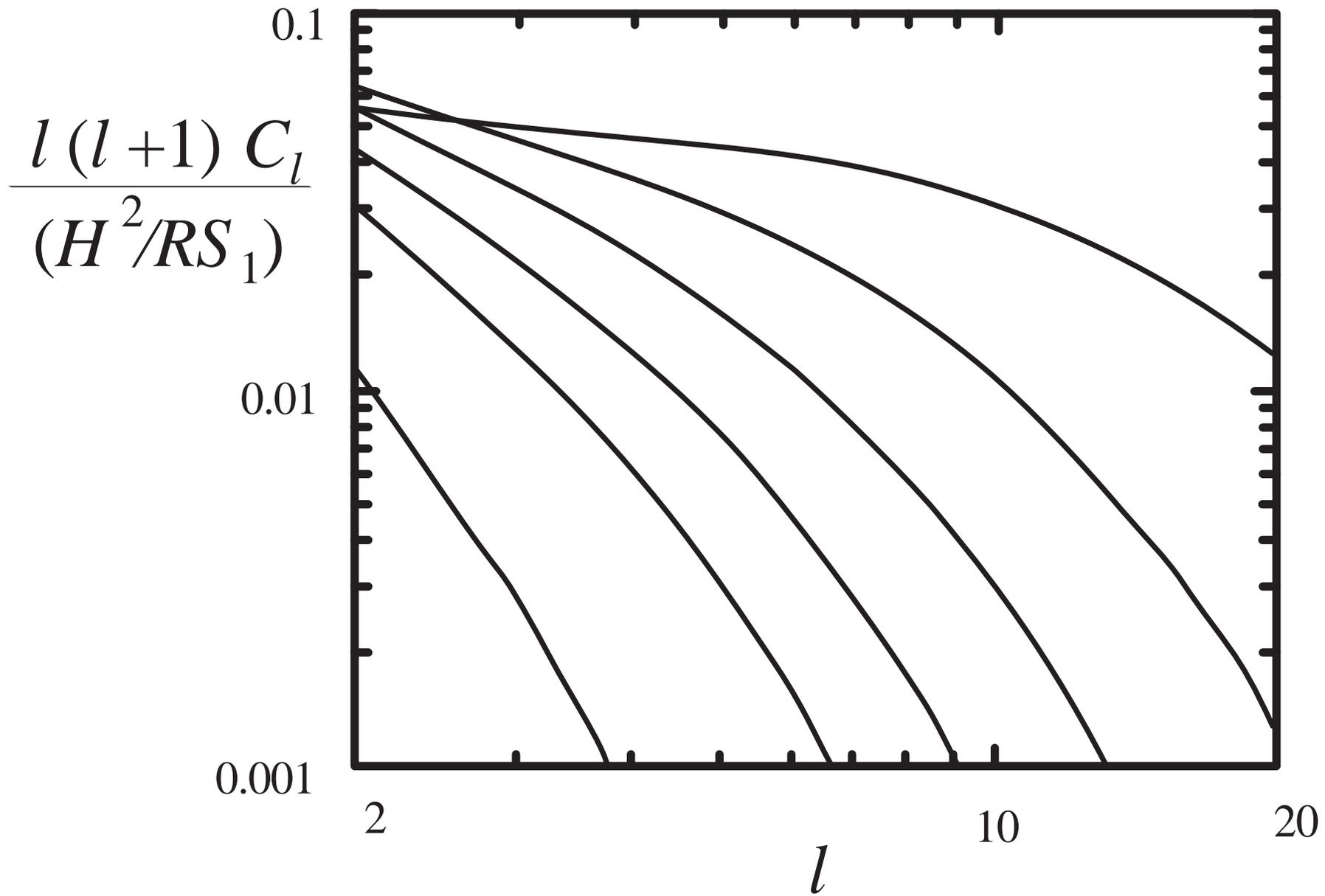
{Figure 4.}
The power spectrum of CMB anisotropies due to the wall fluctuation mode
for various values of $\Omega_0$. The lines are normalized
by $H^2/(S_1R)$.

\vspace{3mm}
\noindent
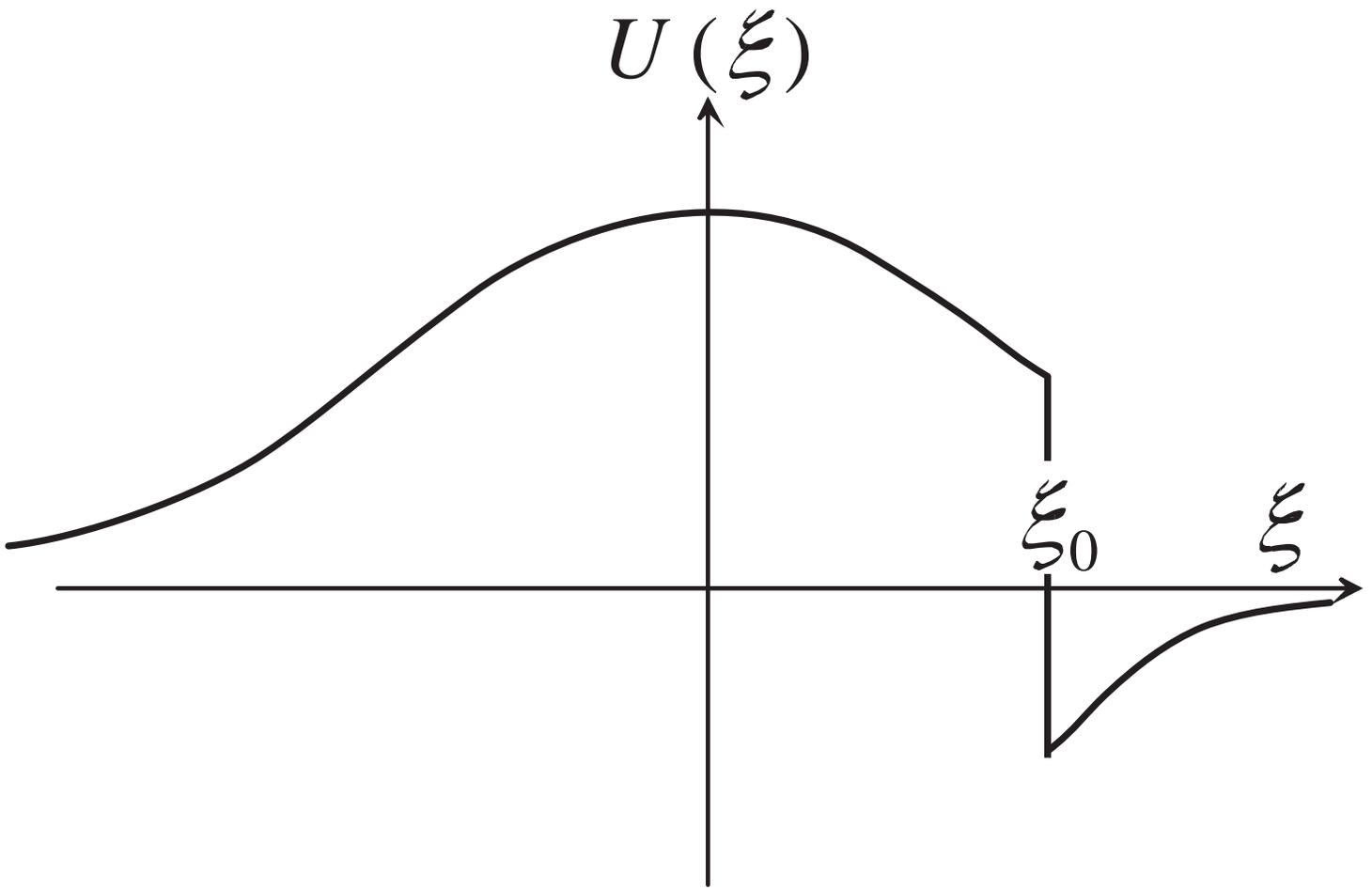
{Figure 5.}
A schematic picture of $U(\xi)$ in eq.(\ref{Uxi})
for a two-field model with thin-wall approximation.

\vspace{3mm}
\noindent
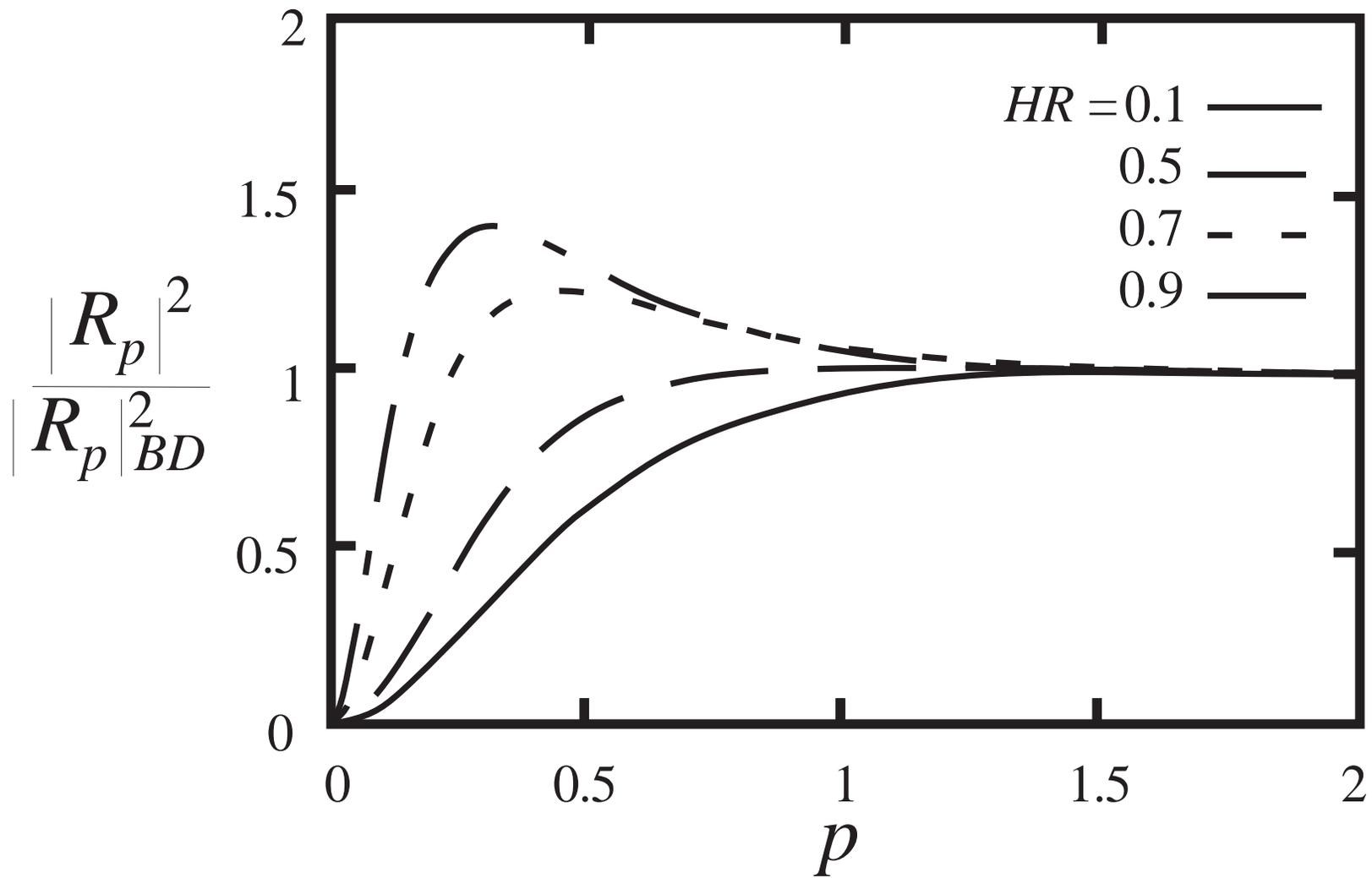
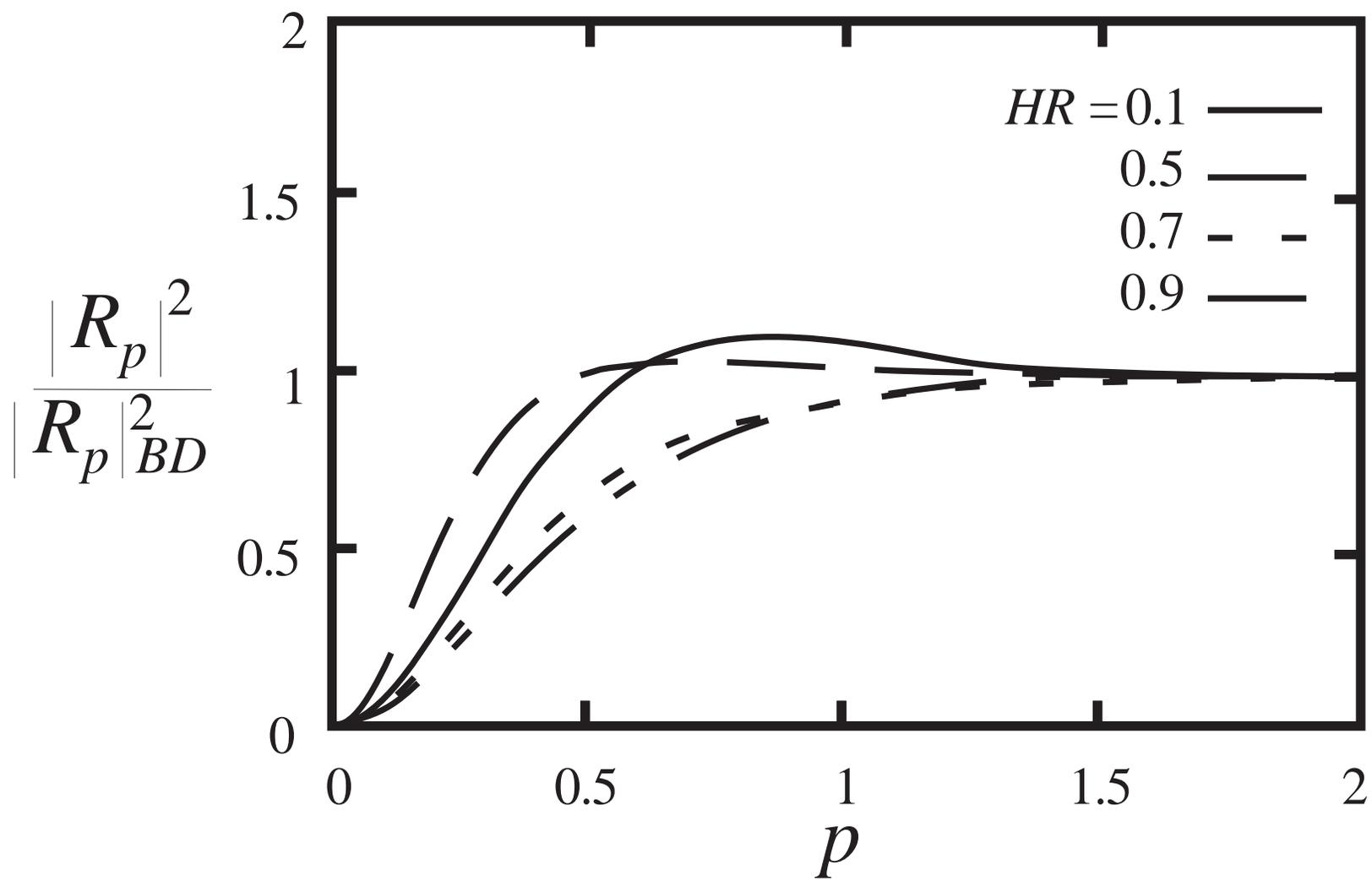
{Figure 6.}
The power spectra of curvature perturbations due to the continuous modes.
The real line, the long dashed line, the short dashed line and the
dotted line are, respectively, for $HR=0.1$, $0.5$, $0.7$ and
$0.9$, for (a) $M/H=2$ and (b) $M/H=10$.

\vspace{3mm}
\noindent
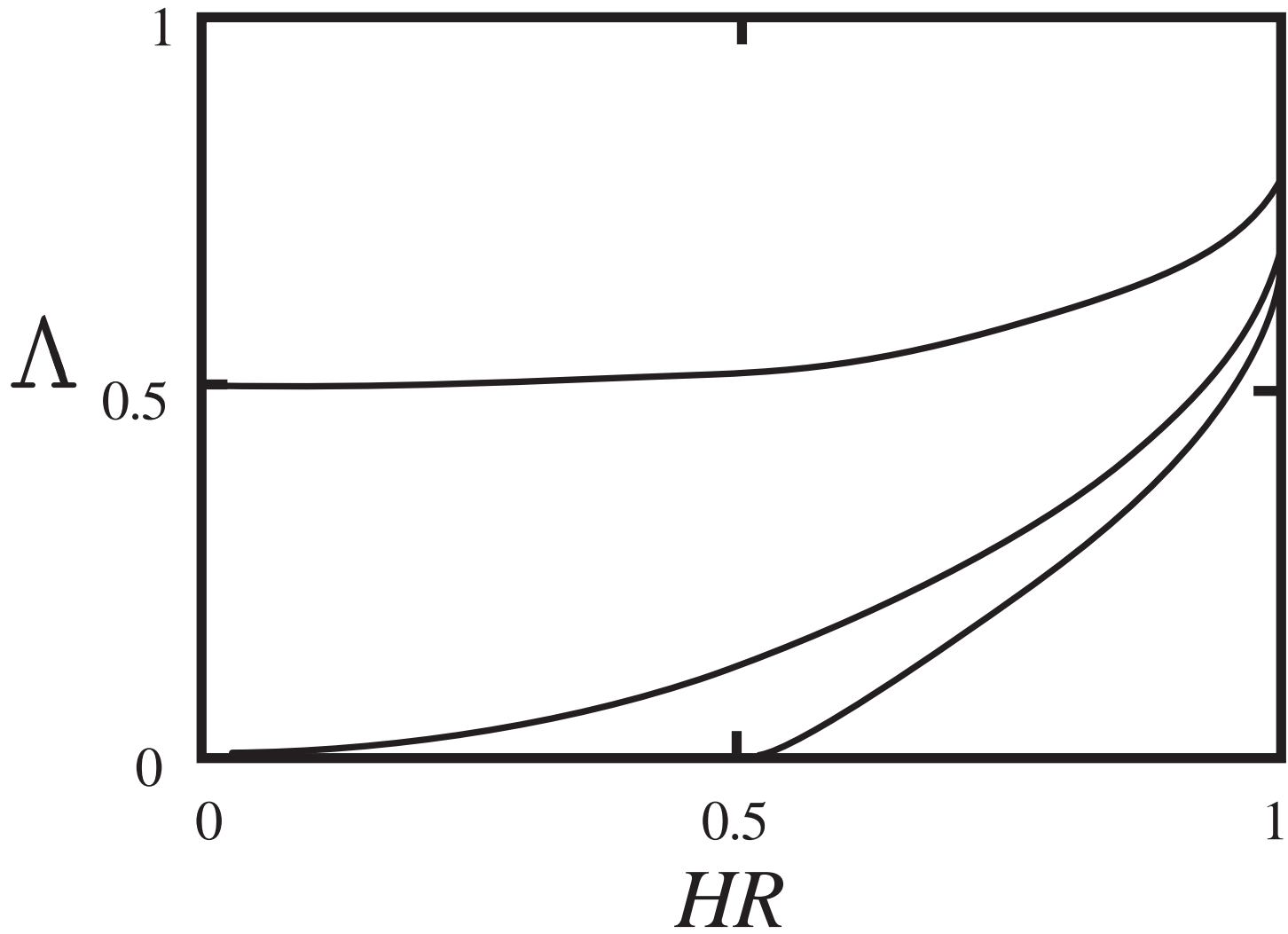
{Figure 7.}
The eigenvalue $\Lambda$ of the de Sitter super-curvature mode
as a function of $HR$ where $R$ is the wall radius.
The three lines, from top to bottom,
 show the cases for $M^2/H^2=5/4$, $2$ and $9/4$, where
$M$ is the mass outside the bubble.

\vspace{3mm}
\noindent
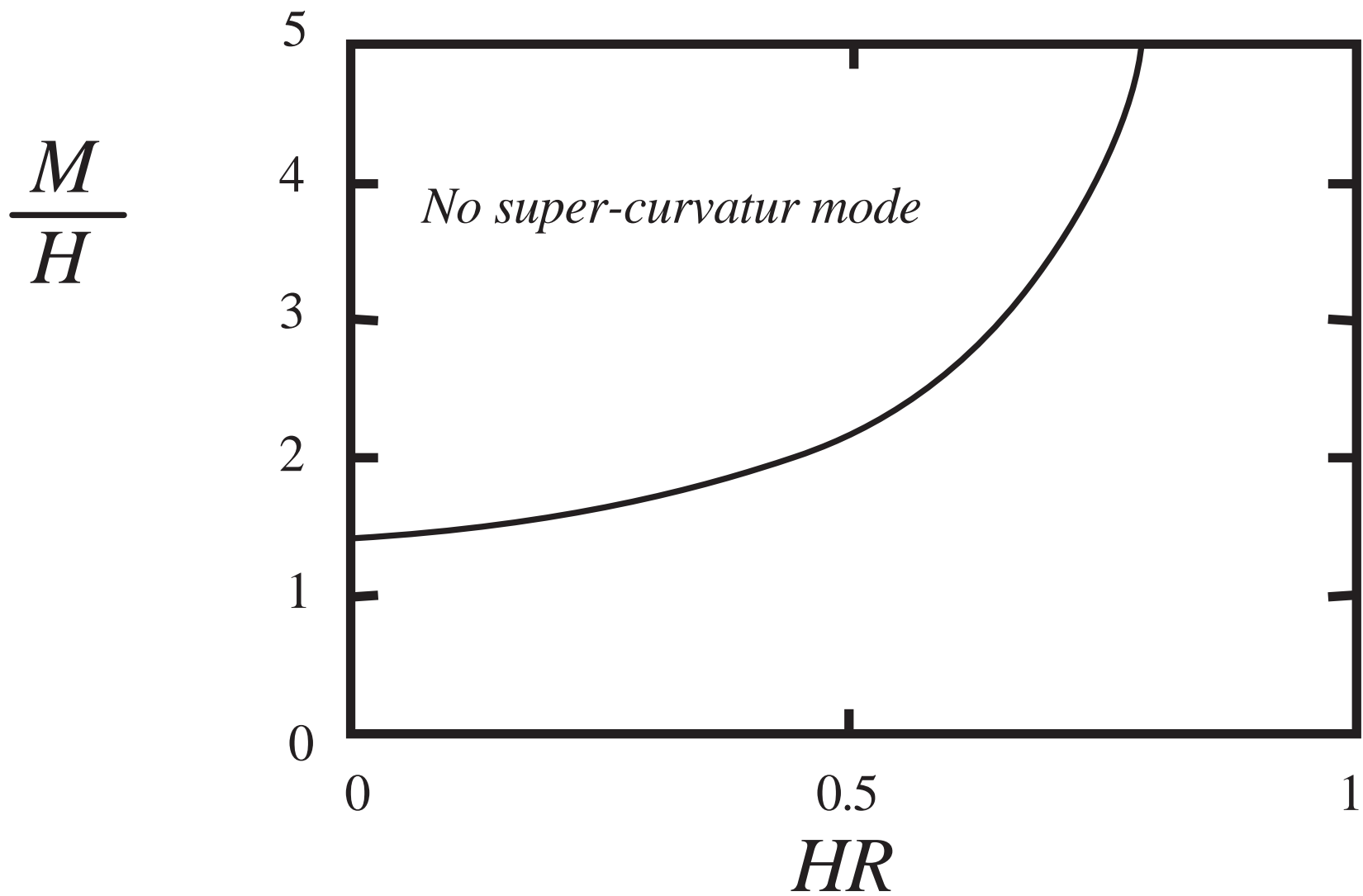
{Figure 8.}
The critical line on which the de Sitter super-curvature mode 
disappears on the $(M/H,HR)$-plane. The super-curvature mode
exists in the region below the line.

\vspace{3mm}
\noindent
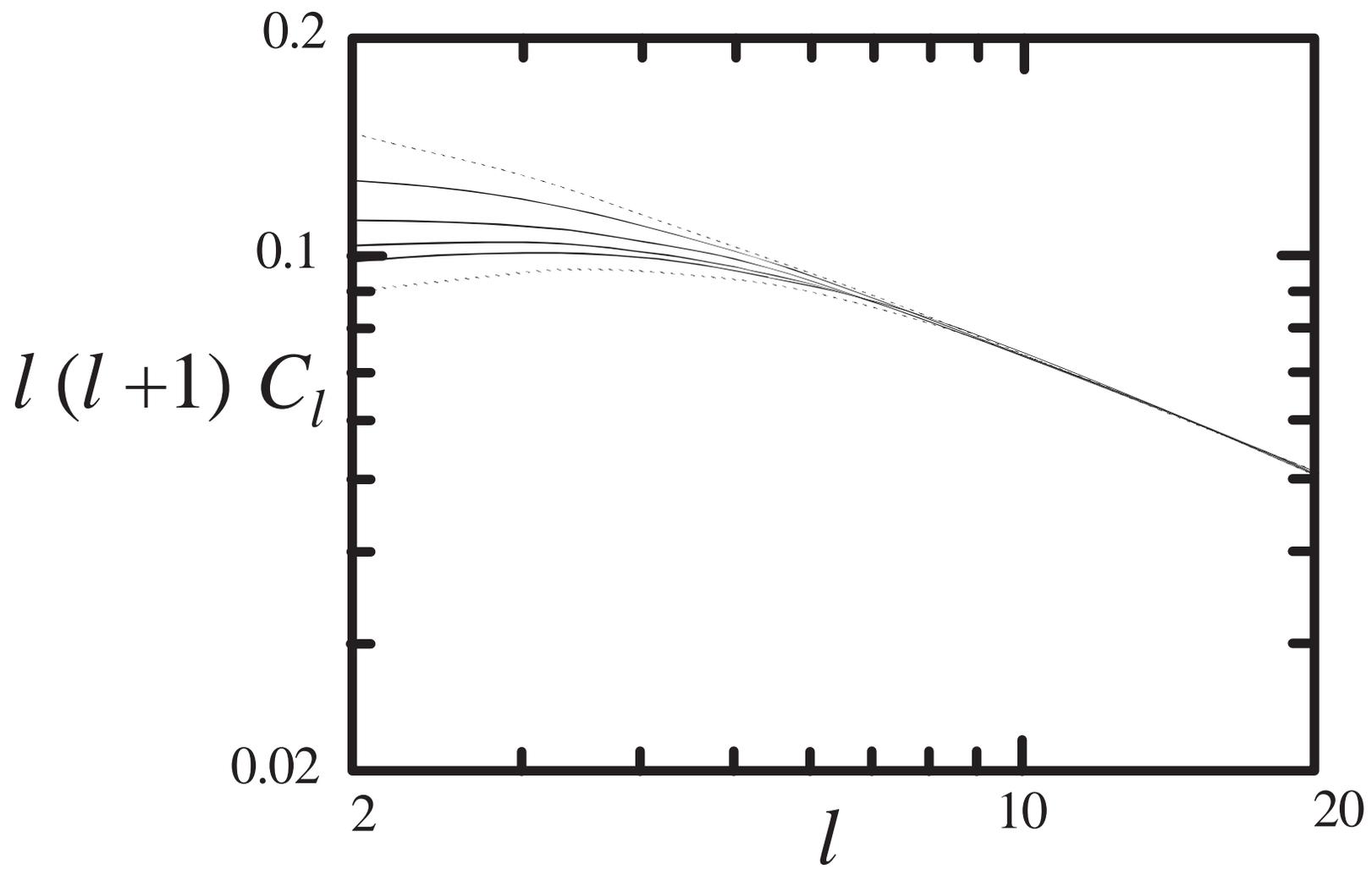
{Figure 9.} 
The CMB anisotropy power spectra in the open universe with
$\Omega_0=0.1$ predicted in two-field models.
 The lines are normalized by $(3H^2/5\dot\phi_B)^2$.
The top and the bottom curves are the results when 
the scalar field is in the Bunch-Davies vacuum and 
the conformal vacuum, respectively.
For the other lines, the model parameters are
 $M^2/H^2=2$ and $HR=0.9$, $0.7$, $0.5$, and $0.3$, 
from top to bottom.

\vspace{3mm}
\noindent
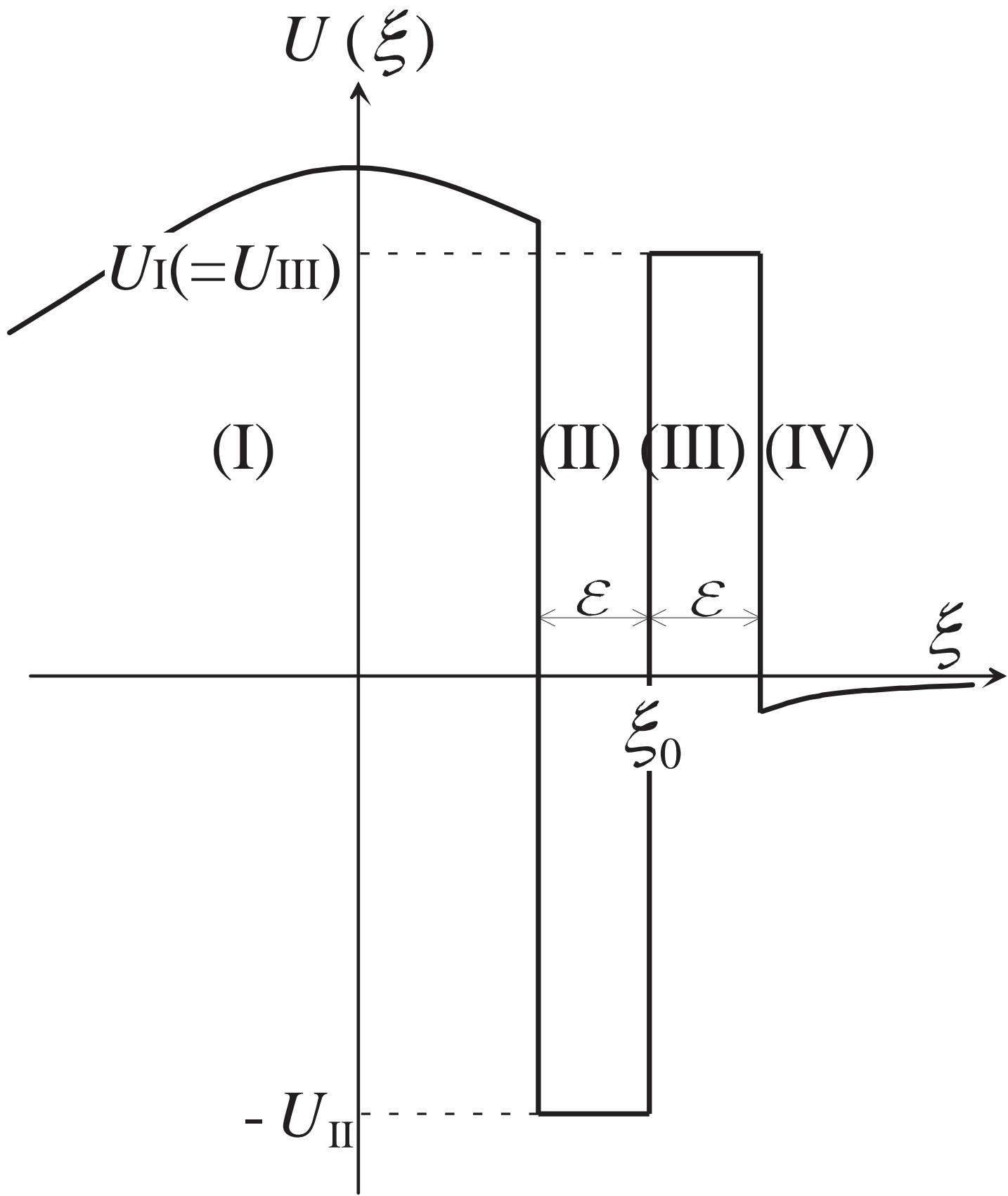
{Figure 10.}
A simplified potential $U(\xi)$ for a single field model.
\end{document}